\documentclass[iop]{emulateapj}
\slugcomment{{\sc Accepted to ApJ:} June 28, 2010} 
\usepackage{amsmath, amssymb}
\usepackage{natbib}
\usepackage{graphicx}

\shorttitle{Dissecting the Red Sequence.  IV.}
\shortauthors{Graves, Faber, \& Schiavon}

\begin{document}

\title{Dissecting the Red Sequence.  IV. The Role of Truncation in the
  Two-Dimensional Family of Early-Type Galaxy Star Formation
  Histories}

\author{Genevieve J. Graves\altaffilmark{1,2,4},
S. M. Faber\altaffilmark{1}, \& Ricardo P. Schiavon\altaffilmark{3}}
\altaffiltext{1}{UCO/Lick Observatory, Department of Astronomy and
  Astrophysics, University of California, Santa Cruz, CA 95064}
\altaffiltext{2}{Department of Astronomy, University of California,
  Berkeley, CA 94720, USA}
\altaffiltext{3}{Gemini Observatory, 670 N. A'ohoku Place, Hilo, HI 96720}
\altaffiltext{4}{Miller Fellow; graves@astro.berkeley.edu}


\begin{abstract}
In the three-dimensional parameter space defined by velocity
dispersion ($\sigma$), effective radius ($R_e$), and effective surface
brightness ($I_e$), early-type galaxies are observed to populate a
two-dimensional fundamental plane (FP) with finite thickness.  In
Paper III of this series, we showed that the thickness of the FP is
predominantly due to variations in the stellar mass surface density
($\Sigma_{\star}$) inside the effective radius $R_e$.  These
variations represent differences in the dark matter fraction inside
$R_e$ (or possibly differences in the initial mass function) from
galaxy to galaxy.  This means that galaxies do not wind up below the
FP at lower surface brightness due to the passive fading of their
stellar populations; they are structurally different.  Here, we show
that these variations in $\Sigma_{\star}$ at fixed dynamical mass
($M_{dyn}$) are linked to differences in the galaxy stellar
populations, and therefore to differences in their star formation
histories.  We demonstrate that the ensemble of stellar population and
$\Sigma_{\star}$ variations through the FP thickness can be explained
by a model in which early-type galaxies at fixed $M_{dyn}$ have their
star formation truncated at different times.  The thickness of the FP
can therefore be interpreted as a sequence of truncation times.
Galaxies below the FP have earlier truncation times for a given
$M_{dyn}$, resulting in lower $\Sigma_{\star}$, older ages, lower
metallicities in both [Fe/H] and [Mg/H], and higher [Mg/Fe].  We show
that this model is quantitatively consistent with simple expectations
for chemical enrichment in galaxies.  We also present fitting
functions for luminosity-weighted age, [Fe/H], [Mg/H], and [Mg/Fe] as
functions of the FP parameters $\sigma$, $R_e$, and $I_e$.  These
provide a new tool for estimating the stellar population properties of
quiescent early-type galaxies for which high-quality spectra are not
available.
\end{abstract}

\section{Introduction}\label{vdml:introduction}

In the cold dark matter (CDM) paradigm, galaxies form hierarchically.
However, more massive galaxies are composed of older stellar
populations than are less massive galaxies (so-called ``archeological
down-sizing'', e.g., \citealt{thomas05}).  Modern semi-analytic models
(SAMs) for galaxy formation have shown that archeological down-sizing
can be produced within the current $\Lambda$CDM framework of
concordance cosmology (e.g., \citealt{delucia06, cattaneo08,
  somerville08, wang08}).  The key to reconciling archeological
down-sizing with hierarchical assembly is to recognize that the first
process refers to the formation of stars, while the second process
refers to the mass build-up of a galaxy, and that these two processes
can be decoupled.  At the same time, galaxies are an orderly family
and obey numerous scaling relations between their stellar populations
and structural properties.  The tightness of the relations indicates
that star formation in galaxies and the mass assembly process, though
decoupled, cannot be combined at random.

This is particularly true of the quiescent, non-star-forming galaxies
discussed in this series of papers.  To zeroth-order, early-type
galaxies form a one-dimensional (1D) sequence, with their star
formation histories and many of their structural properties showing
strong trends with galaxy mass.  However, they are not a purely 1D
family: early-type galaxies populate the two-dimensional Fundamental
Plane \citep[FP]{djorgovski87, dressler87} in the three-dimensional
parameter space of galaxy properties defined by central velocity
dispersion ($\sigma$), effective radius ($R_e$), and effective surface
brightness ($I_e\equiv L / 2 \pi R_e^2$, where $L$ is galaxy
luminosity).  In fact, the FP has finite thickness, which adds a third
dimension to the early-type galaxy structural parameter space.  Thus
the 1D nature of many galaxy scaling relations disguises a
multi-dimensional galaxy family.

While the high-mass FP appears nearly unchanged since $z \sim 1$ other
than expected passive evolution in $M_{\star}/L$, low-mass galaxies
are still ``settling'' onto the FP since $z \sim 1$
\citep{vanderwel04, treu05-apjl, treu05, vanderwel05}.  The local FP
should then be expected to include at least some galaxies that are
relatively recent arrivals.  This is supported by the results of
\citet{forbes98} and \citet{terlevich02}, who find that residuals from
the FP correlate with age such that galaxies offset to higher (lower)
surface brightness have younger (older) ages than those occupying the
FP, based on simple stellar population (SSP, i.e., single burst)
models.  The thickness of the FP may therefore be an age sequence,
with the most recent arrivals lying at higher surface brightness than
the midplane.  This thickness reveals a (narrow) third dimension to
the structural parameters of early-type galaxies.

This paper is the fourth installment in a series of papers that seeks
to connect the star formation histories of early-type galaxies to
their global structural parameters.  We explore these relations in an
explicitly multi-dimensional parameter space, which allows us to
identify the fundamental trends in a model-free fashion.

It has been well-documented that the stellar population properties of
early-type galaxies scale with their stellar mass ($M_{\star}$) or
with $\sigma$ (e.g., \citealt{kuntschner98, trager00b, terlevich02,
  proctor02, thomas05, nelan05, gallazzi05, smith07, graves07}).
However, just as the Faber-Jackson relation \citep{faber76} turns out
to be a 1D projection of the 2D Fundamental Plane, we showed in Papers
I and II of this series \citep{graves09_paperI, graves09_paperII} that
the above-mentioned familiar stellar population trends {\it are in
  fact 1D projections of an underlying 2D family of stellar population
  properties}.

The main dimension consists of correlations between stellar population
properties and $\sigma$ (e.g., \citealt{proctor02, thomas05,
  nelan05}).  One of our main findings in Papers I and II was to
demonstrate that in fact the star formation histories of galaxies
depend most strongly on the velocity dispersion $\sigma$ and {\it not}
on stellar mass, dynamical mass, or luminosity.  Thus the various
``mass-dependent'' relations are in reality $\sigma$-dependent
relations.  This conclusion has been corroborated by the results of
\citet{vanderwel09} and \citet{rogers10}.  In addition to this first
dimension of variation, numerous authors have reported a spread in
stellar population ages and metallicties at fixed $\sigma$, with an
anti-correlation observed between age and metallicity (e.g.,
\citealt{worthey95, colless99, jorgensen99a, trager00b, kuntschner01,
  smith07-iaus245}).  In Papers I and II, we not only confirm the
existence of this second dimension of variation, we also show that it
is strongly correlated with other global galaxy properties.

Paper I mapped the 2D family of stellar population properties onto a
familiar 1D early-type galaxy scaling relation: the color-magnitude
relation (CMR).  We showed that the various stellar population
properties behave differently with respect to $\sigma$ and to
magnitude, such that age and [Mg/Fe] show stronger trends with
$\sigma$, while [Fe/H] and [Mg/H] show stronger trends as functions of
magnitude.  A 3D mapping of stellar population properties onto
$\sigma$, color, and magnitude demonstrated that the CMR hides a 2D
family of stellar population properties.  The first dimension of this
family maps most strongly onto $\sigma$, while the second dimension
correlates with magnitude residuals at fixed $\sigma$.\footnote{In a
  subsequent analysis, \citet{smith09_parameters} find a similar 2D
  family of stellar population variations in a sample of galaxies from
  the Shapley Supercluster.  However, they interpret these as due to
  aperture effects in the observations.  This interpretation is
  inconsistent with the results demonstrated in Paper II, which
  strongly support our original interpretation that the stellar
  population effects are intrinsic and genuine.}

Paper II carried the analysis to the standard 3D space of global
galaxy properties: the Fundamental Plane space defined by $\sigma$,
$R_e$, and $I_e$.  This set of parameters is our preferred space in
which to explore stellar population trends because of two important
points shown in Paper II: (1) that on the FP midplane, galaxy star
formation histories are {\it independent of} $R_e$ and therefore
depend explicitly on $\sigma$, not on total galaxy mass, and (2) that
the second dimension of stellar population variation correlates
strongly with surface brightness residuals at fixed $\sigma$ and
$R_e$.  We showed that sorting galaxies by surface brightness
residuals, instead of $L$ residuals (as in Paper I), gives a cleaner
differentiation of the various stellar population properties, and thus
that it is $\Delta I_e$, not $\Delta L$, that drives the second
dimension of variation.  This means that the second stellar population
dimension maps onto {\it a cross-section through the FP}.  This result
was new.  It demonstrated not only that the thickness of the FP is
real, but that it is an important parameter in the evolutionary
history of galaxies.  Furthermore, these trends exist for galaxies at
{\it fixed physical size $R_e$}, which makes it impossible that the
trends could be due to aperture effects alone as claimed by
\citet{smith09_parameters}.  Taken together, Papers I and II show that
the 2D family of galaxy star formation histories maps smoothly onto 3D
FP space.  This paper quantifies and expands on that work.

In \citet[hereafter Paper III]{graves10a}, we computed the variations
in $M_{\star}/L$ implied by the above 3D mapping in FP space (assuming
a constant initial mass function) and compared these to the measured
dynamical mass-to-light ratios ($M_{dyn}/L$).  We showed that, if the
IMF is constant, $M_{\star}/L$ variations are too small to account for
the observed variations in $M_{dyn}/L$, both along the FP midplane
(the ``tilt'' of the plane) and through the thickness of the plane.
Hence, either the IMF is changing throughout FP space in a systematic
way or the fractional contribution of dark matter relative to stars is
changing.  Focusing on the variations through the thickness of the FP
(i.e., variations with $I_e$ at fixed $\sigma$ and fixed $R_e$), we
considered the implications of IMF variations and showed that the most
natural IMF scenarios predict abundance and line index variations that
are {\it qualitatively at variance with those observed.}  This led us
to tentatively conclude that our estimates of stellar mass are valid
and further that the observed variations in $M_{dyn}/L$ imply real
changes in stellar mass surface density ($\Sigma_{\star}$) through the
plane; galaxies below the FP (at lower $I_e$) must have fewer stars
within $R_e$, not just dimmer stars.  {\it This implies that galaxies
  do not evolve below the FP by passive fading---they are built that
  way.}  We went on to consider ways this might happen and proposed a
toy model in which objects with the same $\sigma$ all start forming
stars at roughly the same time but have different durations of star
formation.  In this picture, galaxies below the plane stop forming
stars sooner (i.e., are ``prematurely truncated'') and end up with
relatively less stellar mass, while galaxies above the FP form stars
longer, have higher overall efficiencies in converting gas into stars,
and wind up with more stellar mass.  We showed that this hypothesis is
{\it qualitatively} consistent with all structural and line-index
trends seen through the FP.

\begin{turnpage}
\begin{deluxetable*}{cr@{ ...}lccccc@{$\pm$}cc@{$\pm$}cc@{$\pm$}cr@{$\pm$}cr@{$\pm$}cr@{$\pm$}cr@{$\pm$}cr@{$\pm$}c}
\tabletypesize{\scriptsize}
\tablecaption{Properties of Galaxy Bins and Stacked
  Spectra\label{bin_tab}}
\tablewidth{0pt}
\tablehead{
\colhead{} &
\multicolumn{2}{c}{} &
\colhead{Median} &
\colhead{Median} &
\colhead{} &
\colhead{} \\
\colhead{$\sigma$ bin} &
\multicolumn{2}{c}{$\Delta \log I_e$ bin} &
\colhead{$\log \sigma$} &
\colhead{$\Delta \log I_e$} &
\colhead{$N$\tablenotemark{*}} &
\colhead{$S/N$\tablenotemark{\dag}} &
\multicolumn{2}{c}{H$\beta$} &
\multicolumn{2}{c}{$\langle$Fe$\rangle$} &
\multicolumn{2}{c}{Mg $b$} &
\multicolumn{2}{c}{[O\textsc{iii}]$\lambda$5007\tablenotemark{\ddag}} &
\multicolumn{2}{c}{Age\tablenotemark{\S}} &
\multicolumn{2}{c}{[Fe/H]\tablenotemark{\S}} &
\multicolumn{2}{c}{[Mg/H]\tablenotemark{\S}} &
\multicolumn{2}{c}{[Mg/Fe]\tablenotemark{\S}} \\
\colhead{(km s$^{-1}$)} &
\multicolumn{2}{c}{($L_{\odot}$ pc$^{-2}$)} &
\colhead{(km s$^{-1}$)} &
\colhead{($L_{\odot}$ pc$^{-2}$)} &
\colhead{} &
\colhead{({\AA}$^{-1}$)} &
\multicolumn{2}{c}{({\AA})} &
\multicolumn{2}{c}{({\AA})} &
\multicolumn{2}{c}{({\AA})} &
\multicolumn{2}{c}{({\AA})} &
\multicolumn{2}{c}{(Gyr)} &
\multicolumn{2}{c}{(dex)} &
\multicolumn{2}{c}{(dex)} &
\multicolumn{2}{c}{(dex)} 
}
\startdata
1.86--2.00  &$-0.25$  &$-0.15$   &1.96  &$-0.19$   &\phantom{1}139  &175  &1.91    &0.05   &2.40    &0.07   &3.23    &0.06   &$-0.15$    &0.03    &\phantom{ }8.1    &0.8   &$-0.26$   &0.04   &$-0.16$    &0.05    &0.10    &0.03\\
  &$-0.15$  &$-0.05$   &1.95  &$-0.09$   &\phantom{1}278  &259    &2.11    &0.03   &2.33    &0.05   &3.32    &0.04   &$-0.25$    &0.02    &\phantom{ }5.5    &0.3   &$-0.22$   &0.03   &$-0.04$    &0.04    &0.18    &0.03\\
  &$-0.05$  &\phantom{$-$}0.05   &1.94  &\phantom{$-$}0.00   &\phantom{1}399  &332  &2.16    &0.03   &2.37    &0.04   &3.27    &0.03   &$-0.19$    &0.02    &\phantom{ }5.0    &0.2   &$-0.18$   &0.02   &$-0.03$    &0.03    &0.15    &0.02\\
  &\phantom{$-$}0.05  &\phantom{$-$}0.15   &1.95  &\phantom{$-$}0.10   &\phantom{1}313  &318  &2.20    &0.03   &2.36    &0.04   &3.22    &0.03   &$-0.19$    &0.02    &\phantom{ }4.6    &0.3   &$-0.17$   &0.02   &$-0.03$    &0.03    &0.14    &0.02\\
  &\phantom{$-$}0.15  &\phantom{$-$}0.25   &1.94  &\phantom{$-$}0.18   &\phantom{1}107  &201  &2.40    &0.04   &2.37    &0.06   &3.06    &0.05   &$-0.20$    &0.03    &\phantom{ }3.1    &0.2   &$-0.10$   &0.03   &\phantom{$-$}$0.02$    &0.04    &0.12    &0.03\\
\hline \\[-0.1in]
2.00--2.09  &$-0.25$  &$-0.15$   &2.05  &$-0.18$  &\phantom{1}216  &241  &1.96    &0.04   &2.41    &0.05   &3.42    &0.04   &$-0.25$    &0.02    &\phantom{ }7.3    &0.5   &$-0.22$   &0.03   &$-0.08$    &0.04    &0.14    &0.03\\
  &$-0.15$  &$-0.05$   &2.05  &$-0.09$   &\phantom{1}556  &414  &2.09    &0.02   &2.42    &0.03   &3.50    &0.02   &$-0.28$    &0.01    &\phantom{ }5.5    &0.2   &$-0.16$   &0.02   &\phantom{$-$}$0.02$    &0.03    &0.18    &0.02\\
  &$-0.05$  &\phantom{$-$}0.05   &2.05  &\phantom{$-$}0.00   &\phantom{1}747  &527  &2.10    &0.02   &2.47    &0.02   &3.50    &0.02   &$-0.26$    &0.01    &\phantom{ }5.3    &0.2   &$-0.13$   &0.01   &\phantom{$-$}$0.03$    &0.02    &0.16    &0.01\\
  &\phantom{$-$}0.05  &\phantom{$-$}0.15   &2.06  &\phantom{$-$}0.09   &\phantom{1}415  &427  &2.25    &0.02   &2.51    &0.03   &3.45    &0.02   &$-0.32$    &0.01    &\phantom{ }3.7    &0.1   &$-0.05$   &0.02   &\phantom{$-$}$0.09$    &0.03    &0.14    &0.02\\
  &\phantom{$-$}0.15  &\phantom{$-$}0.25   &2.06  &\phantom{$-$}0.18   &\phantom{1}130  &261  &2.38    &0.03   &2.47    &0.05   &3.38    &0.04   &$-0.29$    &0.02    &\phantom{ }3.1    &0.1   &$-0.02$   &0.03   &\phantom{$-$}$0.13$    &0.03    &0.15    &0.02\\
\hline\\[-0.1in]
2.09--2.18  &$-0.25$  &$-0.15$   &2.13  &$-0.18$   &\phantom{1}304  &303  &1.94    &0.03   &2.48    &0.04   &3.69    &0.03   &$-0.30$    &0.02    &\phantom{ }7.4    &0.4   &$-0.16$   &0.02   &\phantom{$-$}$0.02$    &0.03    &0.18    &0.02\\
  &$-0.15$  &$-0.05$   &2.14  &$-0.09$   &\phantom{1}898  &590  &1.99    &0.02   &2.51    &0.02   &3.84    &0.02   &$-0.29$    &0.01    &\phantom{ }6.7    &0.2   &$-0.11$   &0.01   &\phantom{$-$}$0.10$    &0.02    &0.21    &0.01\\
  &$-0.05$  &\phantom{$-$}0.05   &2.14  &\phantom{$-$}0.00   &1331            &797  &2.03    &0.01   &2.56    &0.02   &3.78    &0.01   &$-0.28$    &0.01    &\phantom{ }5.9    &0.2   &$-0.08$   &0.01   &\phantom{$-$}$0.10$    &0.01    &0.18    &0.01\\
  &\phantom{$-$}0.05  &\phantom{$-$}0.15   &2.14  &\phantom{$-$}0.09   &\phantom{1}838  &696  &2.11    &0.01   &2.57    &0.02   &3.72    &0.01   &$-0.26$    &0.01    &\phantom{ }5.1    &0.1   &$-0.05$   &0.01   &\phantom{$-$}$0.12$    &0.01    &0.17    &0.01\\
  &\phantom{$-$}0.15  &\phantom{$-$}0.25   &2.14  &\phantom{$-$}0.18   &\phantom{1}267  &443  &2.29    &0.02   &2.62    &0.03   &3.53    &0.02   &$-0.32$    &0.01    &\phantom{ }3.4    &0.1   &\phantom{$-$}$0.04$   &0.01   &\phantom{$-$}$0.17$    &0.02    &0.13    &0.01\\
\hline\\[-0.1in]
2.18--2.27  &$-0.25$  &$-0.15$   &2.22  &$-0.18$   &\phantom{1}254  &320  &1.91    &0.03   &2.43    &0.04   &4.06    &0.03   &$-0.31$    &0.02    &\phantom{ }7.7    &0.4   &$-0.17$   &0.02   &\phantom{$-$}$0.13$    &0.03    &0.30    &0.02\\
  &$-0.15$  &$-0.05$   &2.22  &$-0.09$   &\phantom{1}871  &676  &1.84    &0.01   &2.54    &0.02   &4.09    &0.02   &$-0.20$    &0.01    &\phantom{ }8.5    &0.2   &$-0.13$   &0.01   &\phantom{$-$}$0.11$    &0.02    &0.24    &0.02\\
  &$-0.05$  &\phantom{$-$}0.05   &2.23  &\phantom{$-$}0.00   &1435            &960  &1.93    &0.01   &2.62    &0.01   &4.05    &0.01   &$-0.25$    &0.01    &\phantom{ }7.1    &0.1   &$-0.06$   &0.01   &\phantom{$-$}$0.15$    &0.01    &0.21    &0.01\\
  &\phantom{$-$}0.05  &\phantom{$-$}0.15   &2.22  &\phantom{$-$}0.09   &\phantom{1}919  &839  &2.00    &0.01   &2.65    &0.01   &3.96    &0.01   &$-0.23$    &0.01    &\phantom{ }6.1    &0.2   &$-0.02$   &0.01   &\phantom{$-$}$0.16$    &0.01    &0.18    &0.01\\
  &\phantom{$-$}0.15  &\phantom{$-$}0.25   &2.22  &\phantom{$-$}0.18   &\phantom{1}264  &494  &2.09    &0.02   &2.66    &0.02   &3.84    &0.02   &$-0.18$    &0.01    &\phantom{ }5.0    &0.2   &\phantom{$-$}$0.01$   &0.01   &\phantom{$-$}$0.17$    &0.01    &0.16    &0.01\\
\hline\\[-0.1in]
2.27--2.36  &$-0.25$  &$-0.15$   &2.31  &$-0.18$   &\phantom{1}128  &258  &1.78    &0.04   &2.54    &0.05   &4.37    &0.04   &$-0.27$    &0.02    &\phantom{ }9.4    &0.6   &$-0.14$   &0.03   &\phantom{$-$}$0.16$    &0.04    &0.30    &0.03\\
  &$-0.15$  &$-0.05$   &2.31  &$-0.09$   &\phantom{1}569  &636  &1.71    &0.01   &2.61    &0.02   &4.42    &0.02   &$-0.15$    &0.01    &10.5    &0.3   &$-0.13$   &0.01   &\phantom{$-$}$0.15$    &0.02    &0.28    &0.02\\
  &$-0.05$  &\phantom{$-$}0.05   &2.31  &\phantom{$-$}0.00   &\phantom{1}944  &890  &1.79    &0.01   &2.67    &0.01   &4.40    &0.01   &$-0.18$    &0.01    &\phantom{ }8.8    &0.2   &$-0.06$   &0.01   &\phantom{$-$}$0.20$    &0.02    &0.25    &0.02\\
  &\phantom{$-$}0.05  &\phantom{$-$}0.15   &2.31  &\phantom{$-$}0.09   &\phantom{1}617  &781  &1.97    &0.01   &2.72    &0.02   &4.24    &0.01   &$-0.26$    &0.01    &\phantom{ }6.4    &0.2   &\phantom{$-$}$0.03$   &0.01   &\phantom{$-$}$0.24$    &0.01    &0.21    &0.01\\
  &\phantom{$-$}0.15  &\phantom{$-$}0.25   &2.30  &\phantom{$-$}0.18   &\phantom{1}146  &413  &2.09    &0.02   &2.74    &0.03   &4.15    &0.02   &$-0.23$    &0.01    &\phantom{ }4.9    &0.2   &\phantom{$-$}$0.06$   &0.01   &\phantom{$-$}$0.26$    &0.02    &0.20    &0.01\\
\hline\\[-0.1in]
2.36--2.50  &$-0.25$  &$-0.15$   &2.39  &$-0.20$   &\phantom{12}43  &160  &1.68    &0.07   &2.58    &0.09   &4.68    &0.07   &$-0.21$    &0.04    &11.0    &1.3   &$-0.14$   &0.05   &\phantom{$-$}$0.22$    &0.07    &0.36    &0.05\\
  &$-0.15$  &$-0.05$   &2.39  &$-0.09$   &\phantom{1}231  &452  &1.65    &0.02   &2.62    &0.03   &4.71    &0.02   &$-0.11$    &0.01    &11.4    &0.4   &$-0.13$   &0.02   &\phantom{$-$}$0.22$    &0.03    &0.34    &0.02\\
  &$-0.05$  &\phantom{$-$}0.05   &2.39  &\phantom{$-$}0.00   &\phantom{1}377  &646  &1.76    &0.01   &2.77    &0.02   &4.66    &0.02   &$-0.23$    &0.01    &\phantom{ }9.1    &0.3   &\phantom{$-$}$0.01$   &0.01   &\phantom{$-$}$0.27$    &0.01    &0.26    &0.01\\
  &\phantom{$-$}0.05  &\phantom{$-$}0.15   &2.39  &\phantom{$-$}0.08   &\phantom{1}230  &539  &1.85    &0.02   &2.80    &0.02   &4.58    &0.02   &$-0.24$    &0.01    &\phantom{ }7.7    &0.2   &\phantom{$-$}$0.05$   &0.01   &\phantom{$-$}$0.29$    &0.01    &0.24    &0.01\\
  &\phantom{$-$}0.15  &\phantom{$-$}0.25   &2.38  &\phantom{$-$}0.18   &\phantom{12}20  &168  &1.92    &0.05   &2.85    &0.07   &4.26    &0.07   &$-0.18$    &0.03    &\phantom{ }6.7    &0.7   &\phantom{$-$}$0.07$   &0.03   &\phantom{$-$}$0.24$    &0.04    &0.17    &0.03\\
\enddata
\tablecomments{$I_e$ is computed in the $V$-band. \\
{*}{Total number of galaxies in the stacked spectrum.} \\
{\dag}{Effective median $S/N$ of the stacked spectrum.} \\
{\ddag}{[O\textsc{iii}]$\lambda$5007 emission line equivalent widths are
  measured after subtracting off a model for the stellar population
  continuum.  They are expressed as negative values because they
  represent {\it emission}, not {\it absorption}.} \\
{\S}{Stellar population properties are computed using
  emission-corrected values of H$\beta$, assuming $\Delta$H$\beta =
  0.7$ EW([O\textsc{iii}]).}
}
\end{deluxetable*}
\end{turnpage}

This paper develops the theme of trends through the thickness of the
FP still further.  Here we make the relation between stellar
populations and structural variables even more explicit by showing in
detail how the 2D family of stellar population
variations---essentially the ``metallicity hyperplane'' of
\citet{trager00b}---map onto structural slices through the plane.  The
original metallicity hyperplane involves just SSP age and [Fe/H].  To
this we add [Mg/H] and [Mg/Fe] and show how these quantities also vary
on slices through the FP.  We quantify all of these mappings by
computing chi-square fits that express SSP age, [Fe/H], [Mg/H], and
[Mg/Fe] as a function of 3D location in FP space.  Thus the four
classic parameters of old stellar populations can now be predicted
purely from a galaxy's location in FP space.  These ``structural SSP''
quantities can be used to estimate the star formation histories of
galaxies for which high-quality spectra are not available.

Having fit SSP parameters throughout FP space, we are then able to
estimate both the {\it mean epoch of star formation} and the {\it
  duration of star formation} at all points in FP space.  The
additional knowledge of the [Mg/H] and [Mg/Fe] mappings (not
considered in Paper III) proves to be key when interpreting the
observed stellar population variations in terms of star formation
histories.  With these data in hand, we return to the premature
truncation model of Paper III and are now able to make the model {\it
  semi-quantitative} with estimates of how the onset and duration of
star formation actually vary with $\sigma$ and above and below the FP
at fixed $\sigma$.  These results are in good accord with the
premature truncation picture.

This paper is organized as follows.  The data and binning method used
in this analysis are described briefly in section \ref{data}.  In
section \ref{results}, we present the 2D family of galaxy star
formation histories and map it onto a cross-section through the FP.
In section \ref{quant}, we quantify these trends by providing
equations for stellar population age, [Fe/H], [Mg/H], and [Mg/Fe] as
functions of $\sigma$ and $\Delta \log I_e$.  Section
\ref{sfh_section} goes on to intepret the observed stellar population
trends in terms of star formation histories and presents a 2D
schematic model for the past star formation of quiescent, early-type
galaxies.  These are used, in conjunction with the results of Paper
III, to support the premature truncation model.  Section
\ref{discussion} discusses possible physical mechanisms that could
produce premature truncation in some galaxies at fixed $\sigma$, as
well as the effect of dry merging on the trends presented here.
Finally, section \ref{conclusions} summarizes our conclusions.

\begin{figure}[!t]
\includegraphics[width=1.0\linewidth]{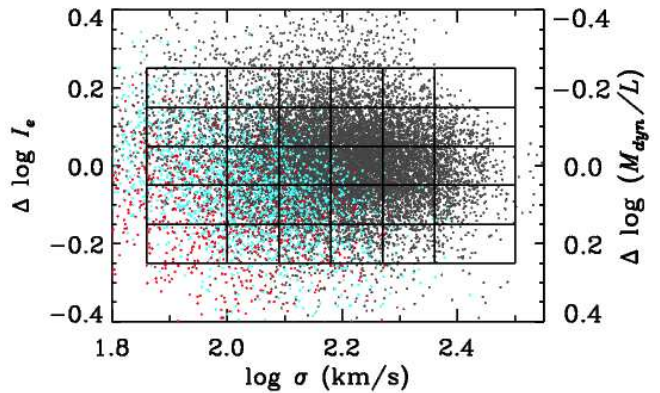}
\caption{Bins used to sort and stack ``similar'' galaxy spectra.  The
  quantity $\Delta \log I_e$ is defined as the difference between the
  measured $\log I_e$ of each galaxy and the best-fitting FP relation
  (see Figure 1 of Paper II).  Because $\Delta \log I_e$ is defined at
  fixed values of $\sigma$ and $R_e$, it is equivalent to $-\Delta
  \log (M_{dyn}/L_V)$.  Colors indicate the absolute magnitude range
  of the galaxies to give a sense of sample completeness.  Dark gray
  points indicate galaxies with $M_r < -20.1$, for which our sample is
  volume-limited.  Red points indicate the faint end of the sample
  ($M_r > -19.7$), where it is missing $\geq 50$\% of the
  volume-limited targets. Cyan points indicate intermediate galaxies
  with $-20.1 < M_r < -19.7$.  Bins with low values of $\sigma$ and
  $\Delta \log I_e$ are substantially incomplete and are populated by
  galaxies from the low-redshift end of our sample ($0.04 < z <
  0.05$).}
\label{bins}
\end{figure}

\section{Data}\label{data}

The galaxy sample used in this work is the same as that used in Papers
I, II, and III.  It consists of $\sim$16,000 early-type galaxies from
the SDSS Main Galaxy Sample \citep{strauss02} in the redshift range
$0.04 < z < 0.08$.  The early-type galaxy selection is described in
detail in Paper I.  Briefly, we require galaxies to have no
significant detectable flux in the H$\alpha$ and
[O\textsc{ii}]$\lambda$3727 emission lines (i.e., line fluxes are both
below a $2\sigma$ detection threshold), to have concentrated light
profiles (i.e., the ratio of the 90\% and 50\% Petrosian radii
$R_{90}/R_{50} > 2.5$ in the $i$-band), and to have light profiles
that are better fit by a de Vaucouleurs profile than by an exponential
profile.  The resultant sample is strongly concentrated on the red
sequence (see Figure 1 of Paper I).  Spectra for the galaxies that
meet these criteria are downloaded from the SDSS Data Archive
Server\footnote{http://das.sdss.org/}.  The various FP parameters are
obtained from the SDSS Data Release 6 (DR6;
\citealt{adelman-mccarthy08}) Catalog Archive Server
(CAS)\footnote{http://cas.sdss.org/dr6/en/} and the NYU Value-Added
Catalog \citep{blanton05-vagc} for Data Release 4 (DR4;
\citealt{adelman-mccarthy06}), as described in Paper II.  The
individual galaxy spectra typically have signal-to-noise $S/N \sim 20$
{\AA}$^{-1}$, so we sort the galaxies into bins based on their FP
parameters, then stack the spectra within each bin to obtain very high
$S/N$ mean spectra that span the binned parameter space.

In Paper II, we showed that the stellar population properties and
therefore the star formation histories of early-type galaxies depend
strongly on $\sigma$ and also on $\Delta I_e$ residuals at fixed
$\sigma$ and fixed $R_e$.  However, on the FP midplane {\it all the
  stellar population properties investigated in Paper II showed no
  dependence on $R_e$}.  In this work we simplify the galaxy
structural space by marginalizing over $R_e$, such that galaxy
properties are defined in the 2D space of $\sigma$ and $\Delta I_e$
only.

\begin{figure*}[!t]
\begin{center}
\includegraphics[width=0.8\linewidth]{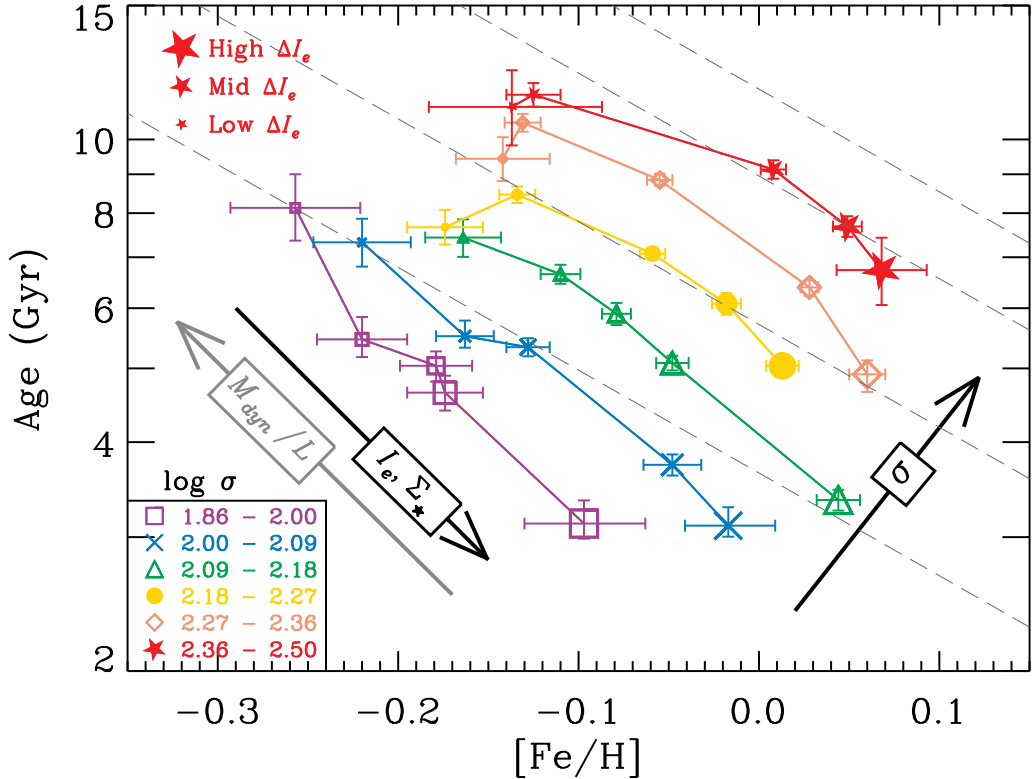}
\end{center}
\caption{The age-[Fe/H] hyperplane for early-type galaxies.  Colors
  and symbols show different bins in $\sigma$, as indicated.  Symbol
  size indicates bins in $\Delta I_e$ at intervals of 0.1 dex, with
  the smallest (largest) symbols indicating the lowest (highest)
  $\Delta I_e$ bins.  Dashed lines are lines of constant $\sigma$ from
  the \citet{trager00b} metallicity hyperplane.  On average,
  high-$\sigma$ galaxies have older ages and higher [Fe/H] than
  low-$\sigma$ galaxies.  At fixed $\sigma$, age and [Fe/H] are
  anti-correlated and this relation is driven by $\Delta I_e$, with
  low (high) $\Delta I_e$ galaxies showing older (younger) ages and
  lower (higher) [Fe/H].  This anti-correlation between age and [Fe/H]
  may indicate that galaxies with low $\Delta I_e$ experienced
  shorter-duration star formation than their high $\Delta I_e$
  counterparts at the same $\sigma$ (see text for details).}
\label{hyperplane}
\end{figure*}

The FP for our galaxy sample is shown edge-on in Figure 1 of Paper II.
Deviations from the plane are parameterized as $\Delta \log I_e$,
where $\Delta \log I_e$ is defined as the difference between the
measured value of $\log I_e$ for each galaxy and the fitted
FP\footnote{Throughout this paper, we will express $\Delta \log I_e$
  as ``$\Delta I_e$'' for convenience, with the understanding that
  this is always computed on a logarithmic scale.}, such that 
\begin{equation}
\Delta I_e \equiv \log (I_e) - (1.16 \log \sigma - 1.21 \log
R_e + 0.55).  \label{fp_plane}
\end{equation}
Note that because $\Delta I_e$ is defined at fixed $\sigma$ and fixed
$R_e$, there is a one-to-one equivalence (although a sign change)
between $\Delta \log I_e$ and $-\Delta \log (M_{dyn}/L)$, since
$M_{dyn}/L \propto \sigma^2 R_e / I_e R_e^2 \propto \sigma^2 / I_e
R_e$.

To group similar galaxies, we define a regular grid in $\Delta
I_e$--$\sigma$ space as shown in Figure \ref{bins}.  The right-hand
axis reflects the equivalence between $\Delta \log I_e$ and $-\Delta
\log (M_{dyn}/L)$.  Within each bin, we stack all galaxy spectra using
a $\sigma$-clipping algorithm and rejecting color outliers, as
described in Paper I.  In each stacked spectrum, we measure the full
set of Lick indices \citep{worthey94,worthey97}, along with estimated
errors derived from the error spectra following the formalism of
\citet{cardiel98}\footnote{All H$\beta$ line strengths have been
  corrected for weak H$\beta$ emission infill, following the procedure
  described in Appendix A of Paper III.}.  We model the mean
luminosity-weighted SSP age, [Fe/H], [Mg/H] and [Mg/Fe], along with
the statistical errors in each quantity, using the stellar population
models of \citet{schiavon07} and the {\it EZ\_Ages} code developed by
\citet[see this work for details of the modeling process]{graves08}.
The properties of the galaxies in each bin, the measured Lick index
values, and the derived stellar population properties are given in
Table \ref{bin_tab}.

As always, when interpreting SSP ``age'' measurements, one should bear
in mind that the quoted age values are {\it Balmer line
  strength-weighted, luminosity-weighted mean ages}.  For populations
with extended star formation histories such as galaxies, the SSP ages
are strongly biased toward the age of the youngest sub-populations of
galaxies (e.g., \citealt{trager09}).  Furthermore, the SSP ages
presented in this paper are averaged over many similar galaxies.
Thus, if one galaxy bin has a younger SSP age than another, this does
not mean that all the stars in all of those galaxies are younger.
Rather, it should be interpreted to mean that at least some (possibly
quite small) fraction of the total star formation in those galaxies
occurred up until more recently in the ``younger'' object than in the
``older'' one.

\section{The 2D Family of Stellar Populations in Early-Type
  Galaxies}\label{results} 

\subsection{The Age-[Fe/H] Hyperplane}\label{hyperplane_section}
  
The first result of this paper is shown in Figure \ref{hyperplane},
which plots the age and [Fe/H] results from the stellar population
analysis for the various bins in $\sigma$ and $\Delta I_e$.  Colors
and symbols give the $\sigma$ values for each bin of galaxies, as
indicated in the figure key.  Symbol sizes indicate bins in $\Delta
I_e$, with the largest and smallest symbols representing the highest
and lowest values of $\Delta I_e$, respectively.  A version of this
figure was presented in Paper III to demonstrate that galaxy star
formation histories vary smoothly through FP space and to discuss
$M_{\star}/L$ modeling.  Here, we present a simplified form, which
collapses all values of $R_e$ into a single point in order to show the
main trends more clearly.  The well-known trends of age and [Fe/H]
with $\sigma$ are recovered: age and [Fe/H] both increase with
increasing $\sigma$.

\begin{figure*}[!t]
\begin{center}
\includegraphics[width=1.0\linewidth]{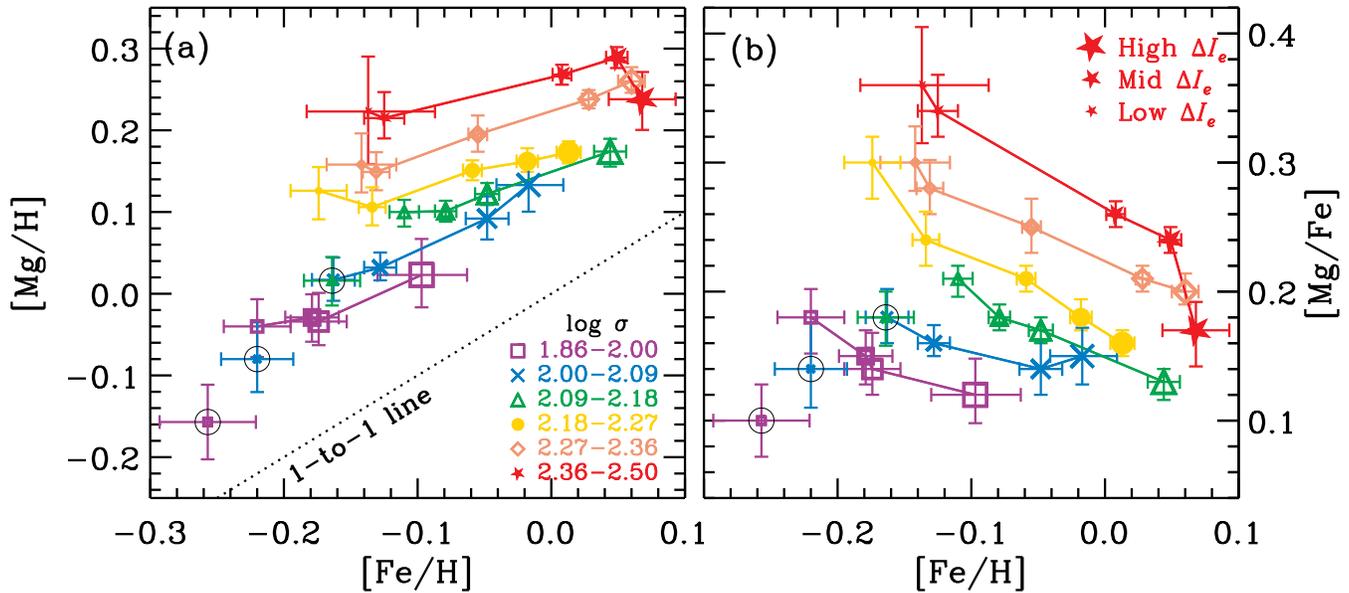}
\end{center}
\caption{(a) [Mg/H] as measured in the stacked spectra, shown as a
  function of [Fe/H].  As in Figure \ref{hyperplane}, colors and
  symbols show different bins in $\sigma$, while symbol size indicates
  bins in $\Delta I_e$. At fixed $\sigma$, both [Fe/H] and [Mg/H]
  increase with increasing $\Delta I_e$.  The slope of the relation is
  shallower than one-to-one (dotted line), indicating that [Fe/H]
  varies more at fixed $\sigma$ than does [Mg/H].  [Mg/H] is always
  higher than [Fe/H], indicating that all galaxies have super-solar
  Mg/Fe ratios, with an offset that increases with $\sigma$.  (b) The
  same plot for [Mg/Fe].  At fixed $\sigma$, [Fe/H] and [Mg/Fe] are
  {\it anticorrelated} with the strongest anticorrelation in the
  highest-$\sigma$ galaxies.  Galaxies with low [Fe/H] span a range of
  [Mg/Fe], but all Fe-rich galaxies have relatively low values of
  [Mg/Fe].  The lowest $\Delta I_e$ bins for the three lowest values
  of $\sigma$ (outlined with black circles) deviate significantly from
  the [Mg/Fe] values expected from the rest of the data, and are
  discussed in detail in section \ref{deviant} }
\label{hyperplane_mg}
\end{figure*}

Furthermore, galaxies within a given $\sigma$ bin span a range in age
and [Fe/H], with age and [Fe/H] anti-correlated at fixed $\sigma$ such
that older galaxies have lower [Fe/H] than their younger counterparts
at the same $\sigma$\footnote{Although the anti-correlation between
  age and [Fe/H] at fixed $\sigma$ is similar to the direction of the
  correlated errors in the stellar population modelling, these
  correlated errors cannot be responsible for the observed trends.
  Firstly, the errors from the modelling (shown in Figure
  \ref{hyperplane}) are too small to contribute substantially to the
  observed spread in the data.  More importantly, if the spread in the
  data were due to random errors in the stellar population modelling,
  it would not be tightly correlated with $\Delta I_e$.  This is
  discussed in detail in Paper I (see section 5.4 and Figure 13) and
  Paper II (see section 4.3 and Figure 11).}.  This anti-correlation
of age and [Fe/H] is essentially identical to the ``metallicity
hyperplane'' of \citet{trager00b} except that our iso-$\sigma$ lines
are slightly converging and curved, whereas those of \citet{trager00b}
were straight and parallel (hence their term hyper{\it plane}).  Lines
of constant $\sigma$ from the Trager et al.\ Fe-hyperplane are
overplotted as dashed lines.  As expected, these run nearly parallel
to our SDSS data for a given value of $\sigma$.  There is a zeropoint
offset between the \citet{trager00b} hyperplane and our iso-$\sigma$
lines due to zeropoint differences between the stellar population
models used in the two analyses.  With the updated models used in this
analysis, we do not find galaxies with ages older than 12 Gyr, whereas
the \citet{trager00b} sample contains galaxies with ages $> 18$ Gyr.

An important result of this series of papers is that the
anti-correlation of age and [Fe/H] at fixed $\sigma$ is driven by
variations in $\Delta I_e$, or equivalently by variations in $\Delta
\log (M_{dyn}/L)$.  This was shown qualitatively in Paper II and will
be quantified below.  The correlation between galaxy star formation
histories and structural properties is such that the oldest, most
Fe-poor galaxies in a given $\sigma$ range are the galaxies with the
lowest values of $\Delta I_e$, while the youngest, most Fe-rich
galaxies are those with the highest values of $\Delta I_e$.
Variations in $\Delta I_e$ are evidently associated with differences
in star formation history between early-type galaxies with the same
$\sigma$.

In Paper III, we demonstrated that stellar population $M_{\star}/L$
effects can only account for a small fraction of the variation in
$I_e$ through the thickness of the FP.  This means that variations in
$I_e$ must be due primarily to variations in the effective stellar
mass surface density in galaxies ($\Sigma_{\star}$), assuming that our
constant-IMF and model stellar mass estimates are valid.  Thus the
oldest, most Fe-poor galaxies at a given value of $\sigma$ are those
with the lowest $\Sigma_{\star}$.

The directions of increasing $\sigma$, $I_e$, $\Sigma_{\star}$, and
$M_{dyn}/L$ are indicated on Figure \ref{hyperplane} by black and gray
arrows, as labelled.  These arrows reveal that the 2D family of
stellar population properties can be mapped onto a cross-section
through the FP, with $\sigma$ along one axis and $\Delta I_e$ along
the other.  In other words, a galaxy's location above or below the FP
is related to how it formed stars, and the metallicity hyperplane
shows how star formation history maps onto slices through FP space.
This mapping is explored further in \S\ref{mapping_section}.

There have been some indications of this variation in previous work.
As mentioned above, \citet{forbes98} and \citet{terlevich02} found
that galaxy ages correlate with residuals from the FP, but in neither
case do they identify or discuss the correlation of FP residuals with
galaxy metallicities.  Simultaneous with our Paper II,
\citet{gargiulo09} reported similar results based on 141 early-type
galaxies in the Shapley supercluster, which show that stellar
population properties, and [Mg/Fe] in particular, correlate with
residuals from the FP.  Their observational results agree with the
results of Paper II, although they parameterize FP residuals in terms
of $R_e$ residuals at fixed $\sigma$ and fixed $I_e$ and therefore
present the correlations differently.

\subsection{Trends in Mg Enrichment}\label{mg_section}

Like [Fe/H], [Mg/H] also shows variation with both $\sigma$ and with
$\Delta I_e$ but the detailed behavior is different.  Figure
\ref{hyperplane_mg}a shows [Mg/H] versus [Fe/H] for the various bins
in $\sigma$ and $\Delta I_e$.  Colors and symbols are as in Figure
\ref{hyperplane} and solid lines connect bins with the same $\sigma$.
At fixed $\sigma$, [Fe/H] and [Mg/H] are correlated with a slope
somewhat shallower than the one-to-one relation (shown as the dotted
line for reference).  The shallowness of the slope indicates that
[Mg/H] changes less at fixed $\sigma$ than does [Fe/H].

The slope of the relation appears somewhat steeper for the lower three
$\sigma$ bins than it does for the higher three $\sigma$ bins.
However, this effect is almost entirely due to three data points: for
the lower three $\sigma$ bins, the lowest $\Delta I_e$ points
(outlined with black circles) have significantly lower values of
[Mg/H] than would be expected from a simple extrapolation of the
relation for the remaining $\Delta I_e$ bins at that $\sigma$.  We
return to these outliers below.

Galaxies at all values of $\sigma$ are offset vertically from the
one-to-one relation (dotted line) such that [Mg/H] is always higher
than [Fe/H].  This indicates that the galaxies in this sample have
super-solar Mg enhancements.  Futhermore, the size of the offset
increases with increasing $\sigma$, indicating that the Mg enhancement
scales with $\sigma$.  This result is by now familiar (e.g.,
\citealt{kuntschner98}, \citealt{jorgensen99a}, \citealt{trager00b}).

Another way to parameterize the Mg-enrichment process in galaxies is
to compare Mg abundances with those of Fe by looking at [Mg/Fe].  This
abundance ratio is not a new independent variable---it is in fact
defined as [Mg/H]$-$[Fe/H]---but it is conceptually useful in the
following discussion to examine Mg trends both in terms of [Mg/H] and
[Mg/Fe].  In Figure \ref{hyperplane_mg}b, we plot [Mg/Fe] as a
function of [Fe/H] for the stacked spectra.  Colors and symbols are as
before.  Galaxies appear to populate a wedge shape in this diagram,
such that all galaxies with high values of [Mg/Fe] have relatively low
values of [Fe/H], while galaxies with low [Mg/Fe] span a range in
[Fe/H].  The galaxies with the highest values of [Mg/Fe] all have
modest values of [Fe/H], suggesting that high Mg-enhancements come at
the cost of high Fe production in chemical evolution.

Previous investigations of the ``metallicity hyperplane'' have not
included [Mg/Fe] among the stellar population properties that vary at
fixed $\sigma$ (e.g., \citealt{trager00b, smith07-iaus245}), while
\citet{thomas05} quantify the intrinsic variation in [$\alpha$/Fe] at
fixed $\sigma$ (they find $\Delta$[$\alpha$/Fe] $\approx$ 0.05 dex)
but do not correlate this scatter with other galaxy observables.  Thus
the addition of [Mg/Fe] to the spread of the metallicity hyperplane at
fixed $\sigma$ is new to this series of papers (see also the
simultaneous work of \citealt{gargiulo09}).

\begin{figure}[!t]
\begin{center}
\includegraphics[width=1.0\linewidth]{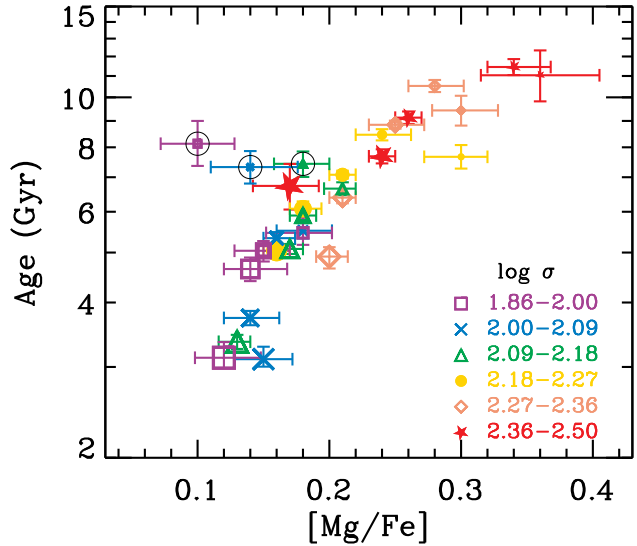}
\end{center}
\caption{Mean light-weighted age versus [Mg/Fe], as measured in the
  stacked spectra.  Colors and symbols are as in Figures
  \ref{hyperplane}--\ref{hyperplane_mg}.  The three outlying data
  points from Figure \ref{hyperplane_mg}b stand out strongly (see
  section \ref{deviant}).  Aside from these, SSP age and [Mg/Fe] are
  strongly correlated, with Mg-enhanced populations having the oldest
  ages.}
\label{mgfe_age}
\end{figure}

In this space, the three outliers identified in Figure
\ref{hyperplane_mg}a stand out even more clearly.  For the three
lowest $\sigma$ bins, the lowest-$\Delta I_e$ galaxies (outlined with
black circles) have unexpectedly low values of [Mg/Fe], which lie as
much as 0.14 dex below the value one would derive from an
extrapolation of the other data.  It is worth noting that, while their
values of [Mg/Fe] are unexpectedly low, their [Fe/H] values are a
reasonable extrapolation of the higher-$\Delta I_e$ data at the same
$\sigma$.  This suggests that it is the Mg enrichment process, not
that of Fe, that deviates in these galaxies.  We return to this point
in section \ref{deviant}.  

Figure \ref{mgfe_age} plots the mean light-weighted ages against
[Mg/Fe] for each of the stacked spectra.  Colors and symbols are as in
Figures \ref{hyperplane}--\ref{hyperplane_mg}.  Ignoring the outlier
data for the moment, there is a tight relation between age and
[Mg/Fe], such that galaxies with the oldest stellar populations are
the most strongly Mg enhanced.  Keeping in mind that the observed SSP
ages are biased toward young subpopulations, this means that galaxies
with some quantity of more recent star formation (i.e., those with SSP
ages $< 6$ Gyr) all have relatively low levels of Mg enhancement
([Mg/Fe] $< 0.2$).  Galaxies with substantial Mg enhancements have
uniformly old populations.

In this projection, the three outlying data points with surprisingly
low [Mg/Fe] stand out very strongly: they do not follow the tight
age-[Mg/Fe] relation of the other galaxies.  The discrepancy varies
systematically with $\sigma$, being largest for the lowest-$\sigma$
bin.  All three points deviate in a consistent way, suggesting that
this deviation is real.  This means that any physical explanation for
the strong age-[Mg/Fe] correlation must also explain why low-$\sigma$,
low-$\Sigma_{\star}$ galaxies might deviate from it.  We will return
to these galaxies in section \ref{deviant}.

\subsection{Mapping the Age-[Fe/H] Hyperplane onto a Cross-Section
  Through the FP}\label{mapping_section}

\begin{figure*}[!t]
\begin{center}
\includegraphics[width=0.8\linewidth]{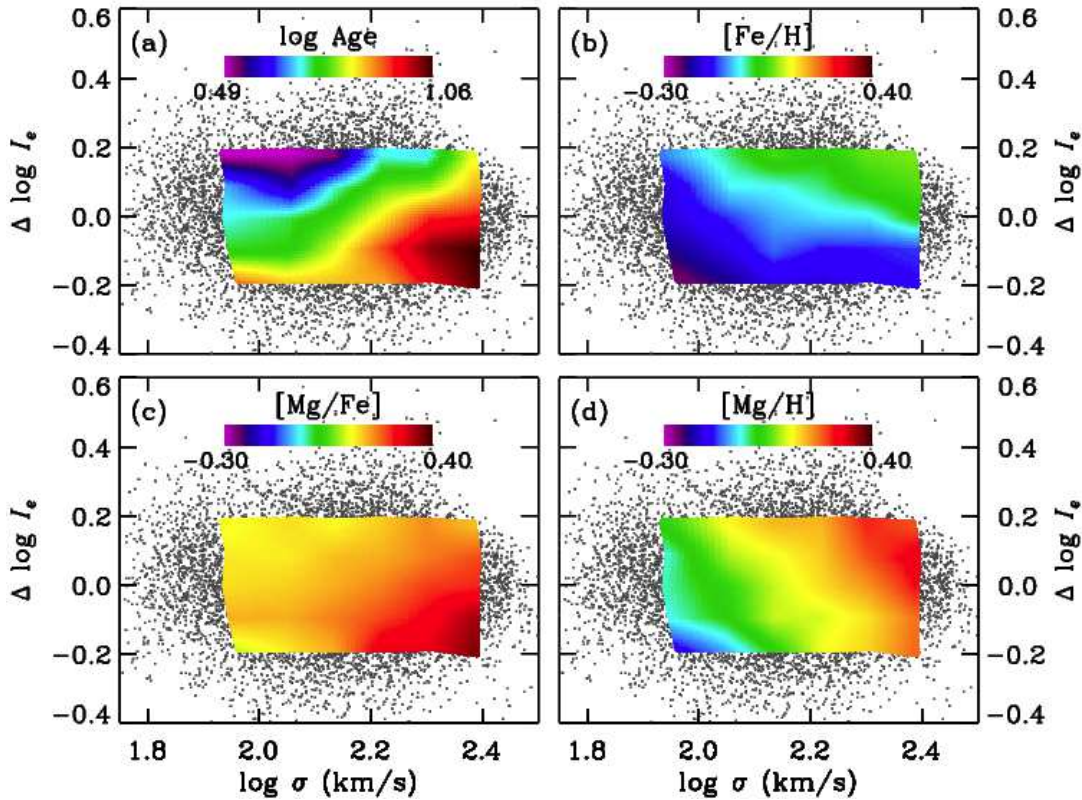}
\end{center}
\caption{Typical stellar population properties of early-type galaxies,
  mapped through $\sigma$--$\Delta I_e$ space.  This space corresponds
  to a cross-section through the FP.  Gray points show the values of
  $\sigma$ and $\Delta I_e$ for individual galaxies in the sample.
  The overlaid color contours show the results of our stellar
  population analysis of the stacked spectra.  Age, [Fe/H], [Mg/H],
  and [Mg/Fe] all increase with $\sigma$.  At fixed $\sigma$, age and
  [Mg/Fe] {\it decrease} with increasing $\Delta I_e$, while [Fe/H]
  and [Mg/H] {\it increase} with increasing $\Delta I_e$.  }
\label{sig_dml_maps}
\end{figure*}

It is clear from Figures \ref{hyperplane}--\ref{mgfe_age} that the
stellar population properties of early-type galaxies depend on both
$\sigma$ and $\Delta I_e$ in systematic ways.  The correspondence
between the structural properties of galaxies and their stellar
populations can be represented by mapping age, [Fe/H], [Mg/H], and
[Mg/Fe] onto $\sigma$--$\Delta I_e$ space.  This mapping is shown in
Figure \ref{sig_dml_maps}.  As defined, $\sigma$--$\Delta I_e$ space
corresponds to a cross-section projected through the FP (i.e.,
marginalizing over $R_e$).  In each panel of Figure
\ref{sig_dml_maps}, gray points indicate the location of the sample
galaxies in $\sigma$--$\Delta I_e$ space, analogous to Figure
\ref{bins}.  The data have been overlaid with color contours
representing the typical values of each stellar population parameter
at that location.  These color contours are computed by plotting the
derived stellar population parameter values for each stacked spectrum
at the point corresponding to the mean values of $\sigma$ and $\Delta
I_e$ for that galaxy bin.  We then interpolate between the $6 \times
5$ grid of values from the stacked spectra to map the stellar
population parameters across the space.

Shown this way, mean luminosity-weighted age (panel a) and [Mg/Fe]
(panel c) track one another fairly closely, as expected from Figure
\ref{mgfe_age}, with lines of constant age and lines of constant
    [Mg/Fe] running diagonally from {\it lower} left to {\it upper}
    right in the maps.  This means that both age and [Mg/Fe] have
    their lowest values for low-$\sigma$, high-$\Delta I_e$ galaxies
    and their maximum values for high-$\sigma$, low-$\Delta I_e$
    galaxies.

In contrast, lines of constant [Fe/H] and lines of constant [Mg/H] run
diagonally from {\it upper} left to {\it lower} right, so that [Fe/H]
and [Mg/H] have their lowest values in low-$\sigma$, low-$\Delta I_e$
galaxies, and their maximum values in high-$\sigma$, high-$\Delta I_e$
galaxies.  This is another illustration that the family of early-type
galaxy stellar population properties is inherently two-dimensional and
further that Fe does not behave like Mg.

Although there is broad agreement in the behavior of age and [Mg/Fe]
and also in the behavior of [Fe/H] and [Mg/H], the trends are
different in detail.  The rough correspondence in the behavior of age
and [Mg/Fe] breaks down at low $\sigma$, where [Mg/Fe] becomes nearly
constant for all galaxies.  This is expected from Figure
\ref{mgfe_age}, which shows a steepening of the age--[Mg/Fe] relation
such that [Mg/Fe] $\sim 0.15$ for a substantial range of ages.  As for
the differences between [Fe/H] and [Mg/H], the contours have different
slopes, such that [Fe/H] shows more variation with $\Delta I_e$ than
does [Mg/H].

\section{Quantifying Stellar Population and Mass-to-Light Variations
  in FP Space}\label{quant}

\begin{figure}[!t]
\begin{center}
\includegraphics[width=1.0\linewidth]{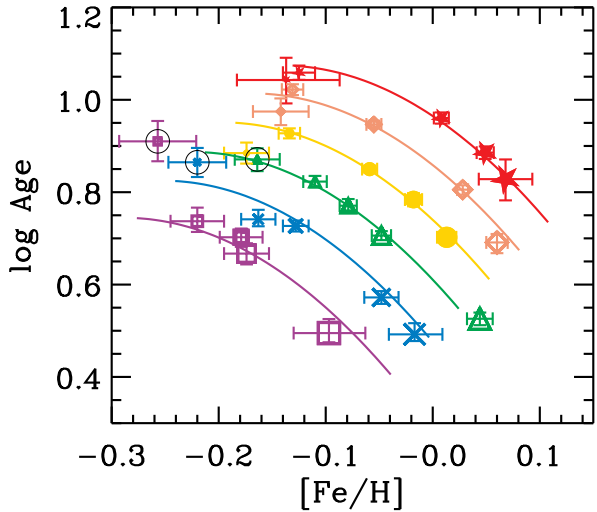}
\end{center}
\caption{Comparing the functional fits to the observed data in
  age-[Fe/H] hyperspace.  The data are as in Figure \ref{hyperplane}.
  The solid lines show the fits from equations \ref{age_fit} and
  \ref{feh_fit}, color-coded to match the corresponding $\sigma$ bins.
  Including a quadratic $(\Delta I_e)^2$ term in the age fit produces
  curvature that closely matches that in the observed galaxies.  Note
  that the lowest-$\Delta I_e$ galaxies in the lower two $\sigma$ bins
  have ages substantially older than those predicted by the fits.
  These are the same galaxies that show unexpectedly low values of
  [Mg/Fe] in Figure \ref{hyperplane_mg}b (see section \ref{deviant}
  for details).  }\label{hyperplane_curves}
\end{figure}

\subsection{Galaxy Properties in 2D: as
  Functions of $\sigma$ and $\Delta
    I_e$}\label{function_section}

The consistent behavior of the various stellar population properties
throughout these FP cross-sections makes it possible to write each
property as a function of $\sigma$ and $\Delta I_e$.  We use the IDL
package {\it mpfit.pro} \citep{markwardt09} to fit robust non-linear
least squares curves in multiple dimensions, taking into account
variable error bars.  A good fit for age as a function of $\sigma$ and
$\Delta I_e$ requires some curvature (as suggested by Figure
\ref{hyperplane}), which we include as a quadratic term in $\Delta
I_e$.  The resulting age relation is
\begin{equation}
\begin{split}
\log age = &\phantom{(}0.82 + \phantom{(}0.75
\phantom{.}\log\sigma' - 0.85
\phantom{.}\Delta I_e - 1.9 \phantom{.}(\Delta I_e)^2\label{age_fit}
\\ &(0.01) \phantom{+} (0.02) \phantom{\log \sigma' -}(0.03)
\phantom{\Delta I_e -}(0.2),
\end{split}
\end{equation}
where $\sigma' = \sigma / 150$ km s$^{-1}$ and $\Delta I_e \equiv \log
I_e - (1.16 \log \sigma - 1.21 \log R_e + 0.55)$, as defined in
Equation \ref{fp_plane}.  The numbers in parentheses give the
1$\sigma$ errors on the quoted parameters, based on the uncertainties
in the measured stellar population parameters (see the values in Table
\ref{bin_tab} and error bars in Figures
\ref{hyperplane}-\ref{hyperplane_mg}).

[Fe/H], [Mg/H], and [Mg/Fe] can all be fit well with simple planar
relations, which give
\begin{equation}
\begin{split}
\mbox{[Fe/H]} = &-0.08 + \phantom{(}0.33
\phantom{.}\log\sigma' + 0.59
\phantom{.}\Delta I_e\label{feh_fit} \\ &\phantom{-}(0.01)\phantom{+}
(0.02) \phantom{\log\sigma'  -}(0.02),
\end{split}
\end{equation}
\begin{equation}
\begin{split}
\mbox{[Mg/H]} = &\phantom{(}0.12 + \phantom{(}0.66
\phantom{.}\log\sigma' + 0.30
\phantom{.}\Delta I_e \label{mgh_fit} \\ &(0.01)\phantom{+}(0.03)\phantom{\log
  \sigma' +}(0.04),
\end{split}
\end{equation}
and
\begin{equation}
\begin{split}
\mbox{[Mg/Fe]} = &\phantom{(}0.20 + \phantom{(}0.28
\phantom{.}\log\sigma' - 0.29
\phantom{.}\Delta I_e\label{mgfe_fit}.\\ &(0.01)\phantom{+}(0.02)\phantom{\log
  \sigma' -}(0.03), 
\end{split}
\end{equation}
respectively.  For [Mg/Fe], the three outlying data points (see
section \ref{mg_section}) were not included in the fit.

Figure \ref{hyperplane_curves} provides a visual confirmation of the
quality of these fits by comparing the age and [Fe/H] fits to the
observed values.  The data, shapes, and color-coding are as in Figure
\ref{hyperplane}.  For each $\sigma$ bin, we compute a curve from the
age and [Fe/H] fits given in equations \ref{age_fit} and
\ref{feh_fit}.  The curve for each bin is computed with $\sigma$ fixed
to the mean value for that bin ($\log \sigma = 1.95$, 2.05, 2.14,
2.22, 2.31, and 2.39 for the six bins, respectively) and for a range
of $\Delta I_e$, with $-0.2 < \Delta \log I_e < 0.2$.  These are
overplotted as solid lines, with colors corresponding to the $\sigma$
bins represented.

The fits do an excellent job of reproducing the data, with the
exception of two points: the lowest-$\Delta I_e$ bins for the two
lower $\sigma$ bins.  In these bins, the observed stellar population
ages are significantly older than expected from the fits.  These are
the same data that show discrepant values of [Mg/Fe] in Figure
\ref{hyperplane_mg}.  They will be discussed in detail in section
\ref{deviant}.

\begin{figure*}[!t]
\begin{center}
\includegraphics[width=0.8\linewidth]{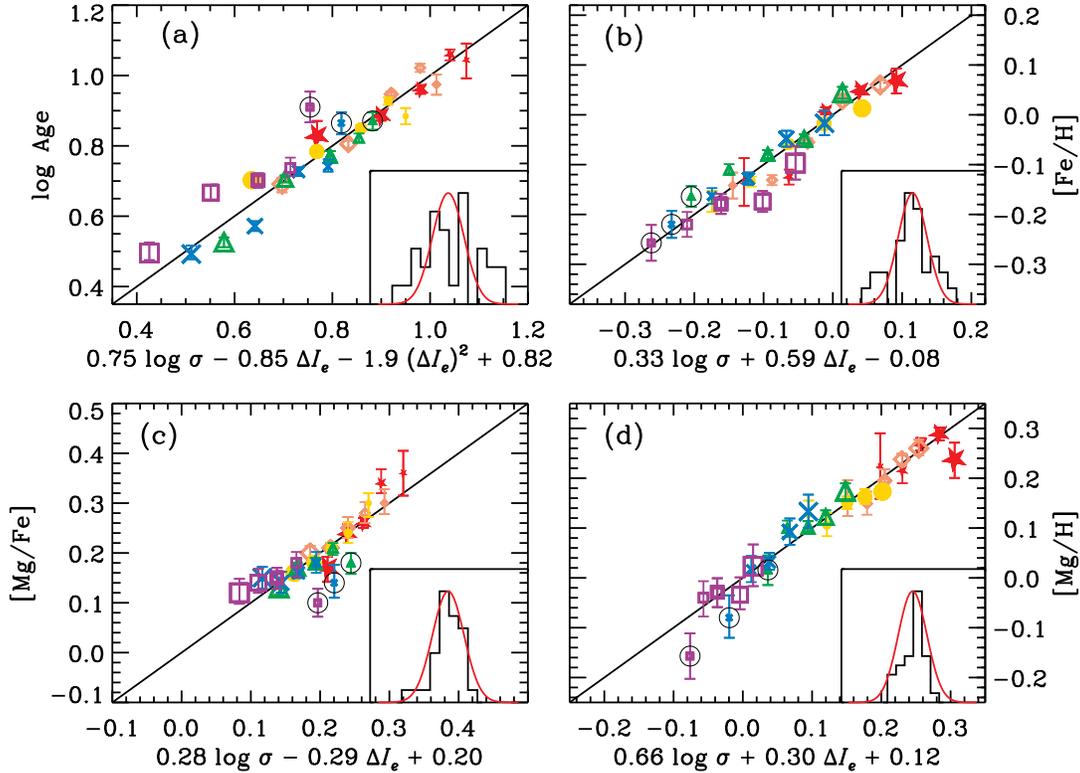}
\end{center}
\caption{Residuals from fits to the stellar population age, [Fe/H],
  [Mg/H], and [Mg/Fe] as functions of $\sigma$ and $\Delta I_e$.  In
  each panel, the x-axis gives the stellar population parameter value
  as predicted by Equations \ref{age_fit}--\ref{mgfe_fit}, based on
  the average value of $\sigma$ and $\Delta I_e$ in each bin of
  galaxies.  The values predicted by the fits are compared to the
  actual values measured from the stacked spectra on the y-axis.
  Solid lines show one-to-one relations.  Colors and symbols represent
  $\sigma$ and $\Delta I_e$ bins, as in Figures
  \ref{hyperplane}--\ref{mgfe_age}.  The outliers are circled in
  black.  The insets show the distribution of offsets from the fits
  (black histogram) in units of the $1\sigma$ measurement errors,
  compared to a gaussian distribution with $\mu = 0$ and $\sigma = 1$
  (over-plotted line).  The insets demonstrate that Equations
  \ref{age_fit}--\ref{mgfe_fit} predict [Mg/H] and [Mg/Fe] within the
  observational errors, but that the difference between predicted and
  observed values for age and [Fe/H] show slightly more scatter than
  can be accounted for by the observational errors.  See text for
  details.}
\label{planar_fits}
\end{figure*}

The quality of the fits is quantified in Figure \ref{planar_fits}.  In
each panel, the x-axis shows the values predicted by Equations
\ref{age_fit}--\ref{mgfe_fit} based on the mean values of $\sigma$ and
$\Delta I_e$ for each galaxy bin, while the y-axis shows the stellar
population parameters as measured in the stacked spectra.  Colors and
symbols are as in Figures \ref{hyperplane}--\ref{mgfe_age}.  When the
stellar population parameters are fit as functions of both $\sigma$
and $\Delta I_e$, the resulting relations are much tighter than would
be derived from fitting just $\sigma$, as can be seen by the
substantial spread in values at fixed $\sigma$.

In each panel, the inset histogram shows the distribution of offsets
from the planar fits expressed in units of the $1\sigma$ measurement
errors for each data point (i.e., the quantity {\it offset/error}).
These can be compared to a gaussian distribution centered at zero with
$\sigma=1$ (overplotted lines).  For [Mg/Fe] and [Mg/H], the fits in
equations \ref{mgh_fit}--\ref{mgfe_fit} replicate the observations
within the observational errors.  The fits to age and [Fe/H] are not
quite as good but still indicate that the fitting functions given in
equations \ref{age_fit} and \ref{feh_fit} are a reasonable
representation of the data.

Overall, the good agreement between the data and the fits illustrates
that the fitting functions give a good representation of the typical
stellar population properties for early-type galaxies as functions of
$\sigma$ and $\Delta I_e$.  They provide a new tool for estimating the
stellar population properties of early-type quiescent galaxies for
which values of $\sigma$, $R_e$, and $L$ can be measured but for which
high-quality spectra are not available.

\subsection{Galaxy Properties in 3D: as
  Functions of $\sigma$, $R_e$, and $\Delta
    I_e$}\label{function_3d_section}

This paper focuses on stellar population variations in the 2D
projected cross-section through the FP.  We have justified dropping
the third FP dimension, $R_e$, because Paper II showed that the
stellar population properties of our quiescent galaxy sample are
independent of $R_e$ at fixed $\sigma$.  It is however instructive to
return to 3D FP space briefly, which we do in this section, in order
to verify (and quantify) the lack of stellar population variation with
$R_e$.  Furthermore, the various mass-to-light ratios, stellar masses,
and stellar mass surface densities of our sample galaxies {\it do}
depend on $R_e$, and we need to be in 3D space to model these
additional quantities.

To do this, we bin the galaxies in a 3D parameter space defined by
$\sigma$, $R_e$, and $\Delta I_e$.  This binning system is described
in detail in Paper II, in which we determine the mean SSP age, [Fe/H],
[Mg/H], and [Mg/Fe] for each bin of galaxies in the 3D space.  In
Paper III, we further measure dynamical and stellar masses ($M_{dyn}$
and $M_{\star,IMF}$), as well as mass-to-light ratios and stellar mass
surface densities using this system.  

Here, we fit each of the stellar population parameters ($\log age$,
[Fe/H], [Mg/H], and [Mg/Fe]) as functions of $\sigma$, $R_e$, and
$\Delta I_e$.  The procedure is similar to that in the previous
section except that here we are fitting in four dimensions instead of
three.  Again, we restrict all fits to be first order in each
parameter with the exception of age, which we allow to include a
quadratic term in $\Delta I_e$.  These fits are summarized in the
first section of Table \ref{fit_table_3d}, labelled ``stellar
populations.''  The quoted errors include only the formal
uncertainties in the fits as estimated by {\it mpfit.pro}.  Presented
in this way, the $\log \sigma$ column of the table indicates how much
the parameter in each row (e.g., $\log age$ or [Fe/H]) changes for
each 1.0 dex change in $\sigma$.  It also means that each row can be
converted into a functional form.  For example, the first row of the
table can be rewritten as an equation for $\log age$, such that
\begin{equation}
\begin{split}
\log age = 0.85 + 0.73 \log \sigma' + 0.02 \log R_e' \\
- 0.92 \Delta I_e - 2.9 (\Delta I_e)^2,
\end{split}
\end{equation}
where $R_e' = R_e / 2.5$ kpc. 

The second section of Table \ref{fit_table_3d} (labelled
``mass-to-light ratios'') gives the parameters of fits for the
dynamical mass-to-light ratio ($M_{dyn}/L$), the component of
$M_{dyn}/L$ contributed by dark matter or IMF variations
($M_{dyn}/M_{\star,IMF}$), and the component of $M_{dyn}/L$
contributed by stellar population variations ($M_{\star,IMF}/L$).  The
third section (labelled ``stellar mass'') also includes the stellar
mass ($M_{\star,IMF}$) and stellar surface mass density within $R_e$
($\Sigma_{\star,IMF}$).  The fits in these two sections are taken from
Table 2 of Paper III, except for the $M_{\star}$ fits, which are newly
computed here in an identical manner to the other fits.  In all cases,
the ``IMF'' subscript indicates a stellar mass or stellar surface mass
density that has been computed {\it assuming a Chabrier IMF}.  The
effect of variable IMFs, if any, would be included in the
$M_{dyn}/M_{\star,IMF}$ term (see Paper III for a thorough discussion
of this point).

Notice that all four stellar population parameters are essentially
independent of $R_e$ in that they show 0.04 dex or less variation per
dex of variation in $R_e$.  The data span almost a full dex in $R_e$,
but no systematic trends in stellar population age, [Fe/H], [Mg/H], or
[Mg/Fe] are observed.  Meanwhile, all four parameters depend strongly
on both $\sigma$ and $\Delta I_e$, with values similar to those from
the 2D fits of section \ref{function_section}.  The differences
between the 3D values in Table \ref{fit_table_3d} and the 2D values in
equations \ref{age_fit}--\ref{mgfe_fit} are in all cases consistent
with the errors quoted for the fits.

\begin{deluxetable*}{lr@{ $\pm$ }lr@{ $\pm$ }lr@{ $\pm$ }lr@{ $\pm$ }lr@{ $\pm$ }l}
\tabletypesize{\footnotesize}
\tablewidth{0pt}
\tablecaption{3D Fits: Stellar Pops and Mass Parameters as Functions of
  $\sigma$, $R_e$, and $\Delta I_e$\label{fit_table_3d}}
\tablehead{
\colhead{} &
\multicolumn{2}{c}{Zeropoint} &
\multicolumn{2}{c}{$\log \sigma$} &
\multicolumn{2}{c}{$\log R_e$} &
\multicolumn{2}{c}{$\Delta I_e$} &
\multicolumn{2}{c}{$(\Delta I_e)^2$}\\
\colhead{Range Covered by Data} &
\multicolumn{2}{c}{} &
\multicolumn{2}{c}{0.5 dex} &
\multicolumn{2}{c}{0.9 dex} &
\multicolumn{2}{c}{0.4 dex} &
\multicolumn{2}{c}{}
}

\startdata
\multicolumn{11}{c}{Stellar Populations} \\
\hline\\[-0.1in]
$\log age$ &$0.85$ &0.01 &$0.73$ &0.05 &$0.02$ &0.03 &$-0.92$ &0.07 &$-2.9$ &0.5 \\
\verb|[|Fe/H\verb|]| &-0.09 &0.01 &$0.27$ &0.04 &$0.04$\tablenotemark{*} &0.02 &$0.64$ &0.04 &\ldots &\ldots \\
\verb|[|Mg/H\verb|]| &0.11 &0.01 &$0.58$ &0.04 &$0.02$\tablenotemark{*} &0.04 &$0.35$ &0.05 &\ldots &\ldots \\
\verb|[|Mg/Fe\verb|]| &0.20 &0.01 &$0.27$ &0.03 &$-0.005$ &0.02 &$-0.27$ &0.04 &\ldots &\ldots \\
\hline \\
\multicolumn{11}{c}{Mass-to-Light Ratios} \\
\hline\\[-0.1in]
$\log \phantom{.}(M_{dyn}/L)$    &0.62 &0.01 &$0.85$ &0.01 &$0.22$ &0.01 &$-1.04$ &0.01 &\ldots &\ldots \\
$\log \phantom{.}(M_{dyn}/M_{\star,IMF})$ &0.20 &0.01 &$0.53$ &0.02 &$0.22$ &0.01 &$-0.80$ &0.02 &\ldots &\ldots \\
$\log \phantom{.}(M_{\star,IMF}/L)$ &0.42 &0.01 &$0.31$ &0.01 &$0.003$ &0.01 &$-0.24$ &0.02 &\ldots &\ldots \\
\hline \\
\multicolumn{11}{c}{Stellar Mass} \\
\hline\\[-0.1in]
$\log M_{\star,IMF}$       &10.6 &0.01 &$1.49$ &0.02 &$0.80$  &0.01 &$0.81$ &0.02 &\ldots &\ldots \\
$\log \Sigma_{\star,IMF}$  &3.81 &0.01 &$1.49$ &0.01 &$-1.20$ &0.01 &$0.80$ &0.02 &\ldots &\ldots \\
\enddata
\tablecomments{$\sigma' \equiv \sigma / 150$ km s$^{-1}$ and $R_e'
  \equiv R_e / 2.5$ kpc.  The units of $I_e$ are $L_{\odot}$ pc
  $^{-2}$.  All $L$ values are $V$-band luminosities.  The ``$IMF$''
  subscript on values of $M_{\star}$ and $\Sigma_{\star}$ indicates
  that they are computed assuming a Chabrier IMF.  Any variation in
  the IMF would be contained within the $M_{dyn}/M_{\star,IMF}$ term.  See
  Paper III for details. \\
{*}{Correcting these values for differences in the
  fraction of the galaxy sampled by the SDSS $3''$ fiber gives the
  values $-0.06$ and $-0.08$ for [Fe/H] and [Mg/H], respectively,
  based on the radial metallicity gradients assumed by
  \citet{smith09_parameters}.  See text for details.} }
\end{deluxetable*}

\citet{smith09_parameters} have suggested that aperture effects may
affect observed trends with $L$ or $R_e$.  The SDSS spectra are
obtained using spectral fibers with diameters of $3''$; the fraction
of the light sampled in each galaxy therefore depends on the galaxy's
size and distance.  If the stellar populations in each galaxy have
substantial radial gradients, these could create spurious (or mask
real) stellar population trends with $R_e$.  Adopting their suggested
metallicity gradients (see section 2.6 of that work) and applying them
to our observed values of [Fe/H] and [Mg/H], we find that the fits in
Table \ref{fit_table_3d} change modestly, such that the $R_e$
parameters for [Fe/H] and [Mg/H] become $-0.06$ and $-0.08$,
respectively (as compared to their uncorrected values of $0.04$ and
$0.02$).  The $\sigma$ and $\Delta I_e$ parameter differences are
within the quoted errors.  These aperture-corrected values still show
very weak dependence on $R_e$, indicating the the stellar population
properties of quiescent galaxies are essentially independent of galaxy
sizes.

As expected from the lack of stellar population variation with $R_e$,
the stellar population component of the mass-to-light variation,
$M_{\star,IMF}/L$, also shows no dependence on $R_e$.  However, the
dark matter (and/or IMF term) {\it does} depend on $R_e$, as do
$M_{\star,IMF}$ and $\Sigma_{\star,IMF}$.  Along the FP, the masses
and mass distributions of galaxies depend on $R_e$ but their star
formation histories do not.

Table \ref{fit_table_3d} illustrates several other interesting points.
It shows the expected inverse relationship between $\Delta I_e$ and
$M_{dyn}/L$ at fixed ($\sigma$, $R_e$) in that $M_{dyn}/L$ changes by
$-1.0$ dex for every one dex change in $\Delta I_e$.  Moreover,
$\sim$80\% of the change in $\Delta I_e$ is due to variations in the
IMF/dark matter component of the galaxy ($M_{dyn}/M_{\star,IMF}$),
with only $\sim$20\% due to stellar population variations
($M_{\star,IMF}/L$).  This is discussed in detail in Paper III.

This dominance of $M_{dyn}/M_{\star,IMF}$ over $M_{\star,IMF}/L$ has a
further implication: most of the change in $\Delta I_e$ is due to a
real change in stellar surface density, $\Sigma_{\star,IMF}$.  This is
also evident from Table \ref{fit_table_3d}, which shows that
$\Sigma_{\star,IMF}$ changes by $\sim$0.8 dex for every 1.0 dex change
in $\Delta I_e$.  Thus the observed $\Delta I_e$ variations about the
FP are predominantly due to differences in the surface mass density of
stars within $R_e$, not to stellar population mass-to-light ratio
effects.  It is therefore not possible for galaxies to evolve through
the FP simply by fading---their stellar mass densities are ab initio
different.  We return to this key point in section \ref{toy_model}.

\subsection{Parameterizing FP Residuals in Terms of $R_e$ versus
  $I_e$}\label{fp_resid_param}

Concurrently with our Paper II, \citet{gargiulo09} published a study
of stellar populations in 3D FP space based on a sample of 141
galaxies in the Shapley Supercluster.  In that work, they parameterize
residuals from the FP in terms of $R_e$ rather than $I_e$.  As in
Paper II, they find that residuals from the FP correlate with stellar
population ages and [Mg/Fe].  The choice of $R_e$ as the dependent
parameter is based upon the fact that the FP was originally proposed
as a tool for measuring distances to galaxies, and therefore the
distance-dependent quantity $R_e$ was historically fit as the free
parameter against the distance-independent quantities $\sigma$ and
$I_e$.

We prefer to parameterize residuals from the FP in terms of $I_e$ for
two reasons.  The first is that comparing galaxies at fixed values of
$\sigma$ and $R_e$ means that they also have the same total dynamical
masses, since $M_{dyn} \propto \sigma^2 R_e$.  More importantly, along
the FP midplane the stellar populations of galaxies are observed to be
independent of $R_e$.  This is true over almost one dex of variation
in $R_e$, which represents a much larger range of variation than the
limited thickness of the FP.  For understanding stellar populations,
it is therefore logical to treat $R_e$ as an ``inert'' variable in the
stellar population modelling and to parameterize the residuals of the
FP and their correlated stellar population variations in terms of
$I_e$.  Of course, for distance studies using $R_e$, it is appropriate
and necessary to set $R_e$ as the dependent variable.

\section{The Star Formation Histories of Early-Type
  Galaxies}\label{sfh_section}

Having fully quantified the mapping between early-type galaxy
structure and stellar populations, let us examine the trends
illustrated in Figures \ref{hyperplane}--\ref{sig_dml_maps} in greater
detail.  The goal is to interpret the observed variations in stellar
population properties as differences in the star formation histories
of the galaxies.

In the following discussion, we will differentiate between two effects
that can modify the abundance patterns of galaxies.  The first is the
effective yield of metals during star formation, due to the chemical
enrichment of the interstellar medium (ISM) by supernovae.  Variations
in effective yield also include differences in the relative effective
yields of different elements, e.g., the Fe-peak elements predominantly
produced by type Ia supernovae (SNe Ia) versus the $\alpha$-elements
predominantly produced by type II supernovae (SNe II).  The effective
yield depends on the IMF, such that top-heavy IMFs with larger
fractions of massive stars produce higher effective yields overall, as
well as higher [$\alpha$/Fe] ratios.  The effective yield also depends
on gas inflow and outflow during star formation; closed box models for
star formation produce different effective yields than those with
substantial outflows of enriched material or inflows of
primordial-abundance gas.  In addition, the effective yield of SNe Ia
elements (e.g., Fe) that is incorporated into stars (and hence is
measured by absorption line studies) depends on the time-scale for
star formation, as discussed below.

A second quantity with relevance for galaxy stellar abundance patterns
is the baryon ``conversion efficiency'' of star formation, which
measures the fraction of gas in a galaxy that is ultimately turned
into stars.  Galaxies with higher conversion efficiencies turn a
larger fraction of their baryons into stars than those with lower
conversion efficiencies.  All other things being equal, this
additional processing of baryonic material results in the production
of more heavy elements over the star-forming lifetime of the galaxy
and should result in higher abundances for all metals.

\subsection{$\sigma$-Dependent Variations in Star Formation
  History}\label{sigma_sfh} 

The various stellar population trends with $\sigma$ have been known
for some time and their implications discussed by many authors.  We
briefly review some of this discussion here before moving on in
section \ref{toy_model} to focus on the new results of this work: the
variations at fixed $\sigma$.

As mentioned above, the effective yield of certain elements may depend
upon the timescale for star formation.  Short timescales for star
formation are often discussed in the literature as a possible
mechanism for producing low-[Fe/H], high-[Mg/Fe] galaxies due to the
time-delay between the production of Mg (and other $\alpha$-elements)
by SNe II and the slower production of Fe by SNe Ia
\citep{tinsley79,greggio83}.  Numerous previous authors (e.g.,
\citealt{worthey92,matteucci94,trager00b,thomas05}) have suggested
that the duration of star formation ($\Delta t_{SF}$) may be shorter
for more massive galaxies, resulting in an increase in [Mg/Fe] with
increasing $\sigma$.  Shorter duration star formation should result in
fewer Fe-peak elements being incorporated into new generations of
stars.  This leads to lower levels of Fe relative to Mg and therefore
to higher [Mg/Fe].  \citet{thomas05} have quantified this using a
simple closed-box chemical evolution model to estimate a quantitative
relationship between [$\alpha$/Fe] and $\Delta t_{SF}$ based on the
chemical evolution models of \citet{thomas99}.

All other galaxy properties being equal, this model for effective
yield variations predicts higher [Mg/Fe] and lower [Fe/H] for galaxies
with shorter $\Delta t_{SF}$.  The predicted trend---an
anti-correlation between [Fe/H] and [Mg/Fe]---is observed {\it at
  fixed $\sigma$} in Figure \ref{hyperplane_mg}.  However, {\it both}
[Fe/H] and [Mg/Fe] are observed to {\it increase} with increasing
$\sigma$.  This suggests that another effect must also be at work
along the $\sigma$-sequence, a different mechanism that modifies the
overall effective yield or changes the conversion efficiency of star
formation versus $\sigma$.  

One oft-cited possibility for this second mechanism involves relaxing
the assumption of a closed-box model to include substantial
quantitites of supernova feedback.  Galaxies with lower values of
$\sigma$ have shallower gravitational potential wells and are
therefore more susceptible to SN feedback, limiting the chemical
enrichment of the interstellar medium by SNe (e.g.,
\citealt{larson74,arimoto87,matteucci94,pipino04}).  This process
lowers the effective yield and tends to produce lower metallicities in
lower-$\sigma$ galaxies.  It should therefore result in lower values
of both [Mg/H] and [Fe/H], as seen in galaxies with low $\sigma$.

If {\it both} of these processes vary with $\sigma$ such that
higher-$\sigma$ galaxies have short $\Delta t_{SF}$ {\it and} are less
affected by SN feedback, it is possible to explain the observed trends
in [Fe/H], [Mg/H], and [Mg/Fe] with $\sigma$.  Changing $\Delta
t_{SF}$ with $\sigma$ can also produced the observed trends in age
(e.g., \citealt{thomas05}), as galaxies which shut down star formation
earlier will have older SSP ages.  This dual picture---overall
modulation by $\sigma$ and variation in $\Delta t_{SF}$ at fixed
$\sigma$---is used to construct a semi-quantitative model in the next
section.

\subsection{A Toy Model for Galaxy Star Formation Histories}\label{toy_model} 

The galaxy data presented in this work also show variations in SSP
age, [Fe/H], [Mg/H], and [Mg/Fe] in a second dimension that correlates
with $\Delta I_e$ at fixed $\sigma$ (e.g., Figure \ref{sig_dml_maps}).
{\it We hypothesize that, in addition to the proposed variations in
  $\Delta t_{SF}$ and SN feedback strength as a function of $\sigma$,
  there are variations in $\Delta t_{SF}$ in early-type galaxy
  progenitors at fixed $\sigma$}.

\begin{figure*}[!t]
\begin{center}
\includegraphics[width=0.8\linewidth]{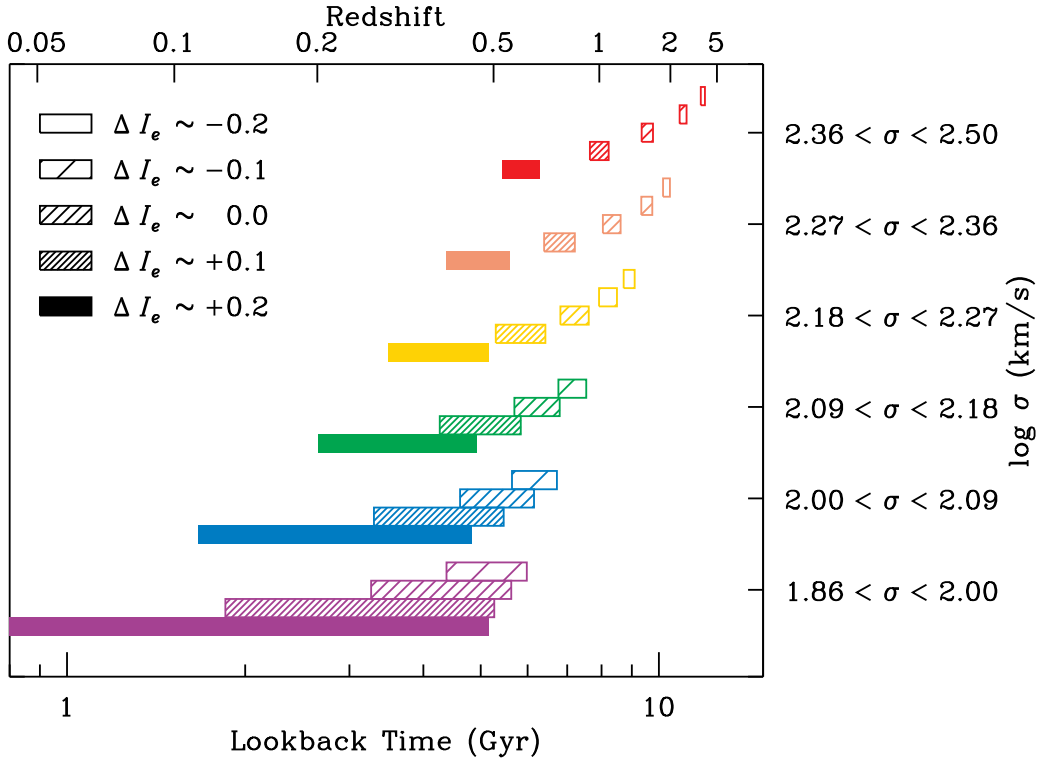}
\end{center}
\caption{Toy models for the star formation histories of early-type
  galaxies, based on the age and [Mg/Fe] fits given in equations
  \ref{age_fit} and \ref{mgfe_fit} (see text for details).  Different
  bins in $\sigma$ are color-coded as in previous figures.  The three
  outlying data bins (see section \ref{mg_section}) are excluded from
  the figure.  Galaxies with low $\sigma$ typically begin star
  formation at later times and form stars over longer timescales than
  galaxies with higher $\sigma$.  At fixed $\sigma$, the duration of
  star formation varies, with the highest-$\Delta I_e$ galaxies
  (filled bars) experiencing the most extended star formation and the
  lowest-$\Delta I_e$ galaxies (empty bars) experiencing the shortest
  duration of star formation.  The range in $\Delta t_{SF}$ at fixed
  $\sigma$ leads to differing mean stellar ages and differing
  enrichment in SN Ia products that are in qualitative agreement with
  the trends shown in Figure \ref{sig_dml_maps}.
}\label{sfh_toymodels}
\end{figure*}

This scenario is illustrated in Figure \ref{sfh_toymodels} by a set of
toy models.  This is a concrete and semi-quantitative representation
of the model we first proposed in Paper III.  In this schematic
representation, star formation is approximated by top-hat models with
constant star formation rates, i.e., star formation is ``on'' where
indicated, and ``off'' everywhere else.  In this simple formulation,
galaxy star formation histories can be represented by two numbers: the
{\it duration} of star formation ($\Delta t_{SF}$) and the {\it onset
  time} at which star formation begins ($t_{on}$).  In the figure,
typical galaxy star formation histories are grouped vertically as a
function of $\sigma$, with color-coding as in Figures
\ref{hyperplane}-\ref{mgfe_age}.  The fill pattern indicates bins in
$\Delta I_e$, as shown in the figure key.  The three outlying data
bins (see section \ref{mg_section}) are excluded from the figure.

The duration of star formation is estimated for each bin by converting
the value of [Mg/Fe] predicted by equation \ref{mgfe_fit} into an
estimate for $\Delta t_{SF}$.  The conversion uses equation 4 from
\citet{thomas05} ([$\alpha$/Fe] $\approx 1/5 - 1/6$ log $\Delta
t_{SF}$).  We substitute [Mg/Fe] for [$\alpha$/Fe] on the assumption
that Mg enrichment traces the group of $\alpha$ elements as a whole.
The Thomas et al. conversion from [Mg/Fe] to $\Delta t_{SF}$ is based
on a closed-box chemical evolution model where star formation
histories are assumed to have a gaussian shape with FWHM $= \Delta
t_{SF}$.  The simplified ``on/off'' representation in Figure
\ref{sfh_toymodels} can be thought of as showing the FWHM peak of star
formation in a gaussian model.

The resulting epoch of star formation is centered on the mean age
predicted for each bin by equation \ref{age_fit}.  Since $\Delta
t_{SF}$ has already been fixed by the value of [Mg/Fe] for each bin,
this effectively sets the value of $t_{on}$.  Of course, these values
are SSP ages rather than a true mass-weighted age.  This means that
the ages as represented in Figure \ref{sfh_toymodels} are skewed
toward young values and later $t_{on}$ than is probably realistic.
However, the goal is not to produce fully realistic, self-consistent
star formation histories\footnote{Producing quantitative and
  self-consistent models would require the development of full
  chemical evolution models, and any quantitative conclusions would
  depend on the detailed shapes of the star formation histories chosen
  for the models.  An intermediate path would be to estimate
  light-weighted ages based on the gaussian star formation histories
  and $\Delta t_{SF}$ values estimated using the \citet{thomas05}
  prescription.  However, this method cannot produce self-consistent
  results, as can be seen from the fact that the predicted values of
  $\Delta t_{SF}$ for galaxies with nearly Solar abundance patterns
  are larger than the age of the universe.\label{delta_tsf_footnote}}
but rather to examine trends for how star formation histories vary
between galaxies with different properties.

If we focus momentarily on ``typical'' galaxies that lie on the FP
midplane (i.e., have $\Delta I_e \sim 0.0$), Figure
\ref{sfh_toymodels} is qualitatively consistent with Figure 10 of
\citet{thomas05}.  Galaxies with higher values of $\sigma$ form stars
over shorter lengths of time and do the bulk of their star formation
at higher redshift than lower-$\sigma$ galaxies.  They follow ``staged
onset'' star formation, in which galaxies with lower values of
$\sigma$ have later $t_{on}$---they begin forming stars at later times
(e.g., \citealt{noeske07b}).  They also manifest ``downsizing''
\citep{cowie99} such that lower-$\sigma$ galaxies continue forming
stars to later times than higher-$\sigma$ galaxies.

The main modification proposed here is the additional variation in
$\Delta t_{SF}$ at fixed $\sigma$, such that galaxies with low $\Delta
I_e$ have shorter duration star formation than those with high $\Delta
I_e$.  In its simplest form, this model would assume that all galaxies
with the same $\sigma$ form the same total $M_{\star}$ but over
different amounts of time (the ``fixed total $M_{\star}$'' model).
This changes the effective yield of Fe, and thus the abundance
patterns of galaxies with the same $\sigma$.  The galaxies with
different $\Delta I_e$ values at fixed $\sigma$ also have slightly
different values of $t_{on}$ such that lower-$\Delta I_e$ galaxies
start forming stars sooner than those with high-$\Delta I_e$.
However, this result depends critically on the quantification of
$\Delta t_{SF}$.

\begin{deluxetable*}{lccccc}
\tabletypesize{\footnotesize}
\tablewidth{0pt}
\tablecaption{Trends in the Toy Models\label{toy_model_table}}
\tablehead{
\colhead{Model} & 
\colhead{$\Sigma_{\star}$} &
\colhead{[Mg/H]} &
\colhead{[Mg/Fe]} &
\colhead{[Fe/H]} & 
\colhead{Age}
}
\startdata
Fixed Total $M_{\star}$ Model   &None   &None   &$+$   &$-$ &$+$ \\
Premature Truncation Model &$-$ &$-$ &$+$ &$-$ &$+$ \\
\hline \\[-0.07in]
Observations &$-$ &$-$ &$+$ &$-$ &$+$ \\ 
\enddata
\tablecomments{Trends give the sign of the effect toward {\it lower}
  $\Delta I_e$.}
\end{deluxetable*}

In general, we caution the reader against drawing strong quantitative
conclusions from this simple toy model, for several reasons.  Firstly,
the quantification of $\Delta t_{SF}$ cannot be correct in detail, as
can be seen by the fact that solar-scale abundances of [Mg/Fe] imply
$\Delta t_{SF} > 15$ Gyr in the prescription of \citet{thomas05}.
Secondly, the SSP values for the stellar population parameters are
skewed from the true mass-weighted values.  This is likely to have a
substantial effect on the stellar ages and a weak effect on the
inferred metallicities and abundance ratios \citep{trager09}.  Thus
the toy model presented in Figure \ref{sfh_toymodels} should be
treated as an illustration of the inferred galaxy star formation
histories, not a strongly quantitative model.  Any quantitative
comparisons with detailed galaxy evolution models should be made
directly with the stellar population data and fitting functions of
section \ref{quant}, rather than with these toy models.

Despite being only semi-quantitative, the toy models serve to
illustrate some general points.  Qualitatively, the star formation
histories in Figure \ref{sfh_toymodels} can explain nearly all of the
trends seen in Figures \ref{hyperplane}--\ref{sig_dml_maps} and Table
\ref{fit_table_3d}, as we now explain.  For galaxies at fixed
$\sigma$, Figure \ref{sig_dml_maps} shows that those with low $\Delta
I_e$ have older ages, higher [Mg/Fe], and lower [Fe/H].  This trend is
consistent with a scenario in which galaxies at fixed $\sigma$ all
begin forming stars at roughly the same time, but experience varying
$\Delta t_{SF}$.  Those galaxies which have short $\Delta t_{SF}$
contain only old stars and have high values of [Mg/Fe] and low values
of [Fe/H] due to the fact that their star formation ceased before the
ISM could fully enrich in Fe-peak elements through SN Ia.  In
contrast, galaxies with longer $\Delta t_{SF}$ form stars until later
times and therefore have younger mean light-weighted ages.  They are
still forming stars as the ISM becomes enriched in Fe, giving their
stellar populations lower values of [Mg/Fe] and higher values of
[Fe/H].

The model star formation histories shown in Figure \ref{sfh_toymodels}
imply that the thickness of the FP represents an age sequence, with
the most recent arrivals on the FP settling onto the high-$\Delta I_e$
layer of the FP.  The youngest ages are seen at low $\sigma$,
suggesting that the most recent additions to the FP are low-mass
galaxies (in agreement with \citealt{treu05}) while the
highest-$\sigma$ galaxies all have ages $> 9$ Gyr.  The ``settling''
onto the high-$\sigma$ end of the FP happened at higher $z$, but those
galaxies also follow the trend that the highest-$\Delta I_e$ galaxies
arrived most recently.

It is tempting to see these correlations between age and $\Delta I_e$
as self-consistent stellar population properties wherein younger
galaxies have higher $I_e$ because they have younger stellar
populations with lower $M_{\star}/L$.  However, as noted in section
\ref{function_3d_section}, this qualitatively plausible reasoning does
not stand up to quantitative analysis.  The stellar population
$M_{\star,IMF}/L$ variations through the FP are small---a by-product
of the anti-correlation between age and metallicity at fixed
($\sigma$, $R_e$).  The observed anti-correlation is in the same sense
as the age-metallicity degeneracy; just as this combination tends to
preserve color, it also tend to preserve $M_{\star,IMF}/L$.  We have
shown in Paper III (as summarized here in Table \ref{fit_table_3d})
that the highest-$\Delta I_e$ galaxies at fixed $\sigma$ and $R_e$
must have considerably higher surface mass densities of stars within
$R_e$ than their low-surface brightness counterparts.  This is at odds
with the simple assumption that all galaxies with the same $\sigma$
form the same total amount of stars.

Thus far we have described the fixed total $M_{\star}$ model, wherein
$\Delta t_{SF}$ varies at fixed $\sigma$ but the same total mass of
stars forms.  This model can explain nearly all of the behavior shown
in Figure \ref{sig_dml_maps}: the increase in age, [Fe/H], [Mg/Fe],
and [Mg/H] with increasing $\sigma$, and the variations in age,
[Fe/H], and [Mg/Fe] at fixed $\sigma$.  But it does {\it not} explain
the {\it absolute variation} in either stellar mass surface density or
[Mg/H] at fixed $\sigma$.  The production and recycling of Mg from
massive stars is nearly instantaneous and therefore the effective
yield of Mg should not change much with variable $\Delta t_{SF}$, as
long as the total amount of star formation remains the same.  In fact,
the total enrichment of [Mg/H] seems to be closely tied to the total
quantity of star formation experienced by a galaxy (Table
\ref{fit_table_3d}).  As discussed in the following sections, the
observed variation in [Mg/H] at fixed $\sigma$ provides valuable
additional information about the galaxy star formation histories
precisely because it is not strongly affected by $\Delta t_{SF}$.

\subsubsection{Premature Truncation
  vs. Fixed Total $M_{\star}$ Models}\label{toy_model_details}

There are two possible simple top-hat models that are consistent with
Figure \ref{sfh_toymodels}.  The first is one in which all galaxies of
the same $\sigma$ form {\it the same total quantity of stars} but
spread out over different amounts of time, so that the galaxies have
different star formation rates.  We have called this the ``fixed total
$M_{\star}$'' model.  The second is one in which all galaxies of the
same $\sigma$ have the same constant star formation rates and
therefore {\it form different quantities of stars over their
  lifetimes} depending on how long star formation lasts.  Galaxies
whose star formation is ``truncated'' earlier will form fewer stars in
total than those that continue star formation to later times.  This
model was introduced in Paper III to fit the trend in $\Sigma_{\star}$
with $\Delta I_e$, where it was termed the ``premature truncation''
model.  These two models represent the extreme possibilities; reality
may lie somewhere in between.

These two models make different predictions for the conversion
efficiency of the star formation process and for the enrichment of
[Mg/H].  A fixed total $M_{\star}$ model is one in which the
efficiency of converting gas into stars is the same for all galaxies
with the same $\sigma$.  Thus the total $M_{\star}$ model predicts
that all galaxies with the same $\sigma$ and $R_e$ will have the same
$M_{\star}$ and the same $\Sigma_{\star}$.  In this scenario, galaxies
should exhibit no variation in $\Sigma_{\star}$ through the thickness
of the FP.  Because they have processed the same quantity of gas into
stars, they should also have the same [Mg/H] (i.e., they should show
no variation in [Mg/H] through the thickness of the FP).

In contrast, the premature truncation model predicts lower conversion
efficiencies for galaxies whose star formation is truncated early.  In
this scenario, the prematurely truncated galaxies (those with lower
$\Delta I_e$) should have lower $M_{\star}$ and lower
$\Sigma_{\star}$.  Furthermore, they have processed less gas into
stars leading to lower production of heavy elements and should
therefore have lower [Mg/H].  In both of these scenarios, shorter
duration star formation leads to higher [Mg/Fe] and lower [Fe/H].

The predicted chemical enrichment pattern for these two versions of
the toy model are summarized in Table \ref{toy_model_table}, along
with the observed variations.  The table shows that the premature
truncation model fits the data better than a fixed total $M_{\star}$
model.  In particular, the observed higher levels of both
$\Sigma_{\star}$ and [Mg/H] in higher $\Delta I_e$ galaxies provide
important leverage for distinguishing between the two normalizations
of the toy model.  The addition of the metallicity and abundance data
here strengthens the case for the premature truncation model proposed
in Paper III, which was based on the observed $\Sigma_{\star}$
variations alone.

\subsubsection{Premature Truncation and a Simple Closed Box Model for
  Enrichment}\label{closed_box} 

We have made the qualitative argument that, at fixed $\sigma$ and
$R_e$ (and therefore fixed $M_{dyn}$), the observed differences in
[Mg/H], $\Sigma_{\star}$, and $M_{\star}$ imply that some galaxies
have proceeded farther in the process of converting gas into stars
than others.  This scenario can be tested in a quantitative way using
a simple model for star formation and chemical enrichment: the closed
box model.

We construct such a model by making several simplifying assumptions.
(1) Galaxies with the same values of $M_{dyn}$ reside in the same mass
dark matter haloes.  (2) The baryon fraction is constant in all haloes.
(3) All baryons start as unenriched gas, which is then processed
through generations of star formation into heavy elements.  (4) We use
the instantaneous recycling approximation, since we are only concerned
with the evolution of Mg enrichment, which is dominated by SNe II.  In
this model, the observed variations in [Mg/H], $\Sigma_{\star}$, and
$M_{\star}$ at fixed $M_{dyn}$ can be directly attributed to the
fraction of gas that has been converted into stars.

\begin{figure}[t]
\includegraphics[width=1.0\linewidth]{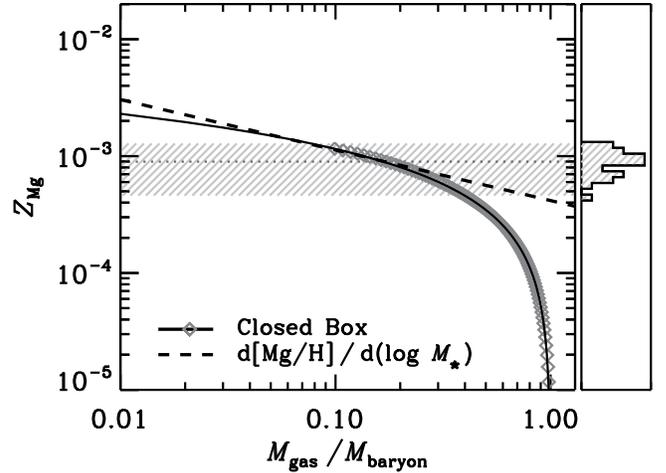}
\caption{Comparing the observed relation between [Mg/H] and
  $M_{\star}$ at fixed $M_{dyn}$ with expectations from a closed box
  chemical enrichment model.  Gray diamonds indicate the closed box
  model, which assumes instantaneous recycling of the SNe II material,
  a Salpeter IMF slope above $M = 8 M_{\odot}$, and the Mg yields of
  \citet{woosley95}.  They show how the Mg-enrichment of the
  interstellar medium evolves as gas is converted into stars (i.e., as
  $M_{\rm{gas}}/M_{\rm{baryon}}$ decreases) and can be represented by
  equation \ref{yield_eq} with $y_{\rm{Mg}} = 0.0005$ (solid black
  line).  The dashed line shows the observed slope of d[Mg/H]/d(log
  $M_{\star}$) for galaxies at fixed $M_{dyn}$ (see text for details).
  The closed box model predicts a similar slope for
  $Z_{\rm{Mg}}$--$M_{\rm{gas}}/M_{\rm{baryon}}$ as the observed
  relation for a wide range of residual gas fractions.  The gray
  shaded region and histogram at the right show the range of
  $Z_{\rm{Mg}}$ spanned by the data.  The observed values of [Mg/H]
  correspond to closed box models with $M_{\rm{gas}}/M_{\rm{baryon}}
  \sim 0.2$ (i.e., 80\% of the gas has been consumed in star
  formation). }\label{closed_box_fig}
\end{figure}

A simple closed box model for chemical evolution gives
\begin{equation}\label{yield_eq}
Z_{\rm{Mg}}(t) = -y_{\rm{Mg}} \ln \left[
  \frac{M_{\rm{gas}}(t)}{M_{\rm{baryon}}} \right],
\end{equation}
where $Z_{\rm{Mg}}(t)$ is the abundance of Mg in the ISM at time $t$,
$y_{\rm{Mg}}$ is the yield of Mg produced by SNe II, and
$M_{\rm{gas}}/M_{\rm{baryon}}$ is the fraction of the initial gas
still remaining in the ISM at time $t$ (\citealt{binney87}, equation
9.17).  We construct a single-zone closed box model and allow stars to
form and enrich the ISM with Mg.  The model uses the Mg yields of
\citet{woosley95}, assuming their higher-energy ``C'' series explosion
models for more massive stars and a Salpeter slope to the IMF above $M
= 8 M_{\odot}$.  This model is shown in Figure \ref{closed_box_fig} as
the gray diamonds and can be fit using equation \ref{yield_eq} with
$y_{\rm{Mg}} = 0.0005$ (solid black line).

This model tracks the relationship between the fraction of gas
consumed in star formation and the Mg-enrichment of the ISM under a
set of simple assumptions.  We can compare it to the slope of the
relation between [Mg/H] and $M_{\star}$ that is observed among the
SDSS galaxies at fixed $M_{dyn}$.  Table \ref{fit_table_3d} shows
that, at fixed $M_{dyn}$ (i.e., fixed $\sigma$ and $R_e$), [Mg/H]
$\propto 0.35$~$\Delta I_e$, while $\log M_{\star} \propto
0.81$~$\Delta I_e$.  Thus the [Mg/H]--$\log M_{\star}$ relation for a
given $M_{dyn}$ should have slope $0.35/0.81 = 0.43$.  This slope is
shown in Figure \ref{closed_box_fig} as the dashed line.  It closely
matches the slope predicted from a closed-box model over a large range
of residual gas fractions.  This indicates that the observed [Mg/H]
variation through the thickness of the FP is at least plausibly
consistent with the enrichment expected from the observed change in
$M_{\star}$.

The shaded region in Figure \ref{closed_box_fig} shows the range of
$Z_{\rm{Mg}}$ spanned by the data, with the dotted line showing the
mean value\footnote{The \citet{anders89} value for the solar Mg
  abundance is used to translate [Mg/H] into $Z_{\rm{Mg}}$.}.  The
distribution of $Z_{\rm{Mg}}$ is indicated by the histogram at the
right.  In the simple closed box scenario, the observed [Mg/H] values
are reached when 10--50\% of the baryons are still in the gas phase,
i.e., when 50--90\% of the gas has been converted into stars.  

Recent studies by \citet{mandelbaum06} and \citet{zaritsky08} find
baryon conversion fractions for early-type galaxies of 12--27\% and
$\sim$10--40\%, respectively.  However, these studies probe the scale
of galaxy groups and clusters.  In these large dark matter haloes, the
closed box model is obviously incorrect as primeval gas continues to
accrete.  At early times while the central galaxy is still forming
stars, some of this gas may make its way onto the galaxy.  This infall
is expected to balance star formation (e.g., \citealt{dekel09}).  With
this type of gas infall, $Z$ rapidly approaches the yield
\citep{binney87} and higher absolute abundances are reached sooner, at
lower $M_{\rm{gas}}/M_{\rm{baryon}}$.  Thus relaxing the assumption of
a closed box should lower the conversion fraction necessary to produce
the observed levels of [Mg/H].  Furthermore, gas accreted at late
times after star formation has ceased will contribute to lower the
baryon conversion fraction for the halo as a whole, as it is not used
in star formation.

Constructing a more realistic chemical evolution model is beyond the
scope of this analysis, but it is interesting to verify that the
observed slope of the [Mg/H]--$\log M_{\star}$ relation at fixed
$M_{dyn}$ is reasonably consistent with expectations from simple
chemical evolution models.

\subsection{Alternative Scenarios}\label{alt_models}

We turn now to other possible scenarios for reproducing the observed
trends in age, element abundance ratios, and stellar mass surface
densities through the FP.  Some of these were also considered in Paper
III.

\subsubsection{IMF Effects}\label{imf}

If the IMF is not constant, appropriate variations might be able to
produce the observed variations in $\Sigma_{\star}$ and $M_{\star}$ at
fixed $\sigma$ and $R_e$.  There are two simple IMF models: (1) a
bottom-heavy IMF that produces more low-mass stars in some galaxies,
leading to under-estimates of $\Sigma_{\star}$ and $M_{\star}$
relative to the true mass, and (2) a top-heavy IMF at early times that
produces more compact remnants at late times, also leading to
under-estimates of $\Sigma_{\star}$ and $M_{\star}$.  As discussed in
Paper III, neither of these models can explain the full set of
observed variations.  A bottom-heavy IMF leaves the relative
frequencies of SNe II and SNe Ia unchanged and therefore does not
produce the observed increase in [Mg/Fe] with decreasing
$\Sigma_{\star}$.  A top-heavy IMF at early times {\it does} increase
[Mg/Fe] as required by the observations.  However, it also produces an
increase in the total production of metals, leading to increases in
[Mg/H] and [Fe/H], opposite to the observed trend (see Table
\ref{toy_model_table}).  Simple models for IMF variation cannot
simultaneously reproduce the observed increase in [Mg/Fe] {\it and}
the decrease in total metal content at low $\Delta I_e$.  

\subsubsection{Differential Dissipation in Galaxy Mergers}\label{dissipation}

In simulations of galaxy mergers, the amount of dissipation changes
the profile of the merger remnant by rearranging the distribution
stars and dark matter with respect to $R_e$ (e.g., \citealt{dekel06,
  robertson06}).  Variations in the degree of dissipation are
therefore another way to produce variations in $\Sigma_{\star}$ and
$M_{\star}$ at fixed $\sigma$.  However, as discussed in Paper III,
the observed trends are not at face value consistent with this
picture.  Highly dissipational mergers should be more common at early
times, when galaxies were in general more gas-rich.  This does not
naturally square with the observation that less dissipational merger
remnants (i.e., galaxies with lower $\Sigma_{\star}$) in fact have
{\it older} stellar populations at a given $\sigma$ (see Table
\ref{toy_model_table}).  It is possible to invoke subsequent
dissipationless merging as a way to increase $M_{dyn}/M_{\star}$ in
systems that form early, but the observed trends extend down to small
galaxies with $\sigma \approx 100$ km s$^{-1}$, where fewer
dissipationless mergers are expected to occur.

\subsubsection{Differential Efficiency of Star
  Formation or Mass-Loss}\label{efficiency}

Other possible models to produce variations in $\Sigma_{\star}$ and
$M_{\star}$ at fixed $\sigma$ and $R_e$ include mechanisms that modify
the efficiency of star formation or of chemical enrichment.

One possibility is that the low-$\Delta I_e$ (low-$\Delta
\Sigma_{\star}$) galaxies form stars for the same duration of time but
overall less efficiently, converting lower fractions of their gas into
stars.  This successfully produces the lower observed values of [Mg/H]
in these objects.  It does not, however, produce their higher observed
values of [Mg/Fe].

Another possibility is that the low-$\Sigma_{\star}$ galaxies form in
lower-mass haloes so that more gas and metals escape in
supernova-driven winds during star formation.  Like the low-efficiency
model above, this model also produces lower $\Sigma_{\star}$ and lower
[Mg/H] but does not produce variation in [Mg/Fe].  

A third possibility is that the low-$\Sigma_{\star}$ galaxies
experience burstier star formation, such that many supernovae go off
at once, producing more violent winds.  This gives lower overall
[Mg/H], as observed, but predicts the wrong trend for [Mg/Fe]; strong
winds during bursty star formation should preferentially remove Mg,
lowering [Mg/Fe].  

It is clear that the combined constraints of {\it lower} [Mg/H] and
{\it higher} [Mg/Fe] in low-$\Sigma_{\star}$ galaxies are powerful
tools for discriminating between galaxy formation models.  None of the
enhanced mass-loss or lower efficiency models discussed in this
section naturally produces both trends simultaneously.

\subsubsection{Summary of Alternative Models}\label{alt_summary}

Faced with meeting both the structural and stellar population
constraints provided here and in Paper III, only the premature
truncation model seems able to match the data without forcing.  The
combined strength of these two types of information proves to have
surprising constraining power and shows the importance of mapping the
stellar population properties into FP space.  Putting these two
different kinds of information together has created a tight test of
galaxy formation models, which at this point seems to leave only one
standing: the premature truncation model.

\subsection{The Outlying Data Points: Low-$\sigma$,
  Low-$I_e$ Galaxies}\label{deviant}

In section \ref{mg_section}, we identified three galaxy bins that
deviate significantly from what one would expect from a smooth
extrapolation of the other data.  These are the lowest-$\Delta I_e$
bins ($-0.25 < \Delta I_e < -0.15$) for each of the three
lowest-$\sigma$ bins ($1.86 < \log \sigma < 2.00$, $2.00 < \log \sigma
< 2.09$, and $2.09 < \log \sigma < 2.18$), which are outlined by black
circles in Figures \ref{hyperplane_mg}, \ref{mgfe_age},
\ref{hyperplane_curves}, and \ref{planar_fits}.  We consider here
several possible explanations for these galaxies.

These galaxies stand out most clearly in Figure \ref{mgfe_age}, where
their [Mg/Fe] values appear {\it much} lower than expected,
particularly for the lowest $\sigma$ bin.  Their age and [Fe/H] values
appear relatively consistent with the other data in Figure
\ref{hyperplane}, but become more discrepant when curvature is allowed
in the age fit of equation \ref{age_fit}, as can be seen in Figure
\ref{hyperplane_curves}.  To summarize, these outliers have ages that
are too old, [Mg/H] abundances that are too low, and [Mg/Fe] values
that are {\it way} too low compared to what one would expect from an
extrapolation of the other data.  The low values of both [Mg/H] and
[Mg/Fe] suggest that these galaxies lost a larger fraction of their
SNe II-enriched material than did their counterparts at the same
$\sigma$ with higher $\Delta I_e$.

One possibility is that these galaxies are substantially affected by
sample selection effects.  They populate the part of galaxy parameter
space that suffers from serious incompleteness (Figure \ref{bins}).
However, is is not clear how this would produce the observed
deviations in stellar population properties.  The galaxy bins contain
enough galaxies to produce adequate S/N in the stacked spectra (Table
\ref{bin_tab}).  They are mostly populated by galaxies from the
lower-redshift ($0.04 < z < 0.05$) portion of our sample.  However,
the quiescent, passive galaxy population we consider here should not
have changed significantly in the $\sim500$ Myr between $z = 0.08$ and
$z = 0.04$ (particularly those at low $\Delta I_e$, which have SSP
ages $> 7$ Gyr), making it difficult to explain the observed
deviations through the effects of evolution within the sample.  The
higher-redshift galaxies in these bins may be erroneously assigned to
the bins due to measurement errors in $\sigma$ or in $\Delta I_e$.
These galaxies should contribute spectra that are representative of
the galaxy properties in other, higher-$\sigma$, higher-$\Delta I_e$
bins.  If the outliers are caused by erroneous bin assignment, they
should represent linear combinations of galaxy properties of other
bins.  This is not the case; these bins have the {\it lowest} observed
values of [Mg/H] and combined [Mg/Fe] and age values that differ
substantially from all other galaxy bins (see Figures
\ref{hyperplane_mg} and \ref{mgfe_age}).

A second possibility is that, if the stellar population properties of
satellite galaxies differ substantially from those of similar galaxies
that are central to their haloes, the observed deviations in these
bins could be due to a higher contribution from satellite galaxies.
We use the group catalogs of \citet{yang07} to identify galaxies which
have the highest $M_{\star}$ in their group and identify these as
central galaxies, while other galaxies are classified as satellites.
The satellite fraction, which is less than one-third in most of the
galaxy bins, increases to a maximum of two-thirds in the
lowest-$\sigma$, lowest-$\Delta I_e$ bin.  However, a preliminary
analysis of the stellar population properties of central galaxies as
compared to satellite galaxies with the same $\sigma$ and $\Delta I_e$
suggests that the two types of galaxies have similar stellar
populations (see also \citealt{cooper09}).  Furthermore, the satellite
fraction of a bin depends more strongly on $\sigma$ than on $\Delta
I_e$; the highest satellite fractions are found in the three bins with
$1.86 < \log \sigma < 2.00$ and $\Delta I_e < +0.05$, rather than the
three bins with $1.86 < \log \sigma < 2.18$ and $\Delta I_e < -0.15$.
We reserve a more detailed analysis of the stellar populations of
satellite versus central galaxies for a future paper, but note here
that our preliminary work on this topic suggests that it cannot
account for the observed outliers.

A third possibility is that the outliers are due to differences in
their susceptibility to feedback.  These galaxies, with the lowest
values of $\Delta I_e$ and low values of $\sigma$, are exactly those
galaxies which we would expect to be most strongly affected by
feedback.  They have the shallowest potential wells (low $\sigma$) and
the least-concentrated mass profiles (low $\Delta I_e$ and low $\Delta
\Sigma_{\star}$).  The outlying galaxies behave in exactly the way
predicted for galaxies that experience substantial mass-loss driven by
SNe II feedback (see section \ref{efficiency}), showing {\it both}
lower than expected values of [Mg/H] and [Mg/Fe].  The most deviant
galaxies are those in the lowest $\sigma$ bin, which would be the most
vulnerable to the removal of Mg-enriched material.  Moreover, the
anomalies increase at lower $\sigma$, futher targeting low $\sigma$
and feedback as the cause.

A problem with this line of argument is that it is not clear why this
effect would kick in suddenly at low $\Delta I_e$ (there is no
indication of deviation for galaxies with similar $\sigma$ and $\Delta
I_e > -0.15$).  This could be explained if, in addition to having the
shallowest gravitational potentials, these galaxies have the burstiest
star formation histories.  The observed low values of [Mg/Fe] in these
galaxies require the preferential loss of Mg-enriched material over
Fe-enriched material.  Multiple short bursts of star formation would
maximize Mg ejection through large numbers of simultaneous SNe II,
while the more extended production of Fe by SNe Ia would be less
efficient at expelling enriched gas, leaving the Fe in the ISM to be
incorporated in the next burst of star formation.  Combining bursty
star formation histories with shallow potential wells could produce
the dramatic transition observed between the main population of
early-type galaxies and these outliers.

\section{Discussion}\label{discussion}

\subsection{Physical Mechanisms for Truncating Star
  Formation}\label{phys_mech} 

The premature truncation model appears to reproduce the complex
ensemble of galaxy data presented in this series of papers.  Multiple
physical mechanisms have been proposed that could provide such a
truncation scenario, which we discuss briefly here.

\subsubsection{Quenching in Infalling
  Satellites}\label{satellite_quenching} 

One possible truncation mechanism for satellite galaxies is the rapid
quenching of star formation when a satellite falls into a massive halo
and is stripped of gas (e.g., \citealt{gunn72, lea76, gisler76}).
This explanation predicts that the prematurely truncated galaxies
(those with lower $\Delta I_e$) at a given $\sigma$ should be
satellites in massive haloes, and that the lowest-$\Delta I_e$
galaxies are those that were accreted into massive haloes at the
earliest times.  There is some evidence that, at fixed $\sigma$,
galaxies in denser environments have slightly older ages than those in
the field (e.g., \citealt{trager00a, thomas05, clemens06, bernardi06,
  smith08, cooper09}).  However, it is not clear that these galaxies
can be associated with the low-$\Delta I_e$ galaxies---they do not
necessarily also show lower metallicities and higher
[Mg/Fe]\footnote{Several authors find no dependence of metallicity or
  abundance pattern on environment at fixed $\sigma$ \citep{thomas05,
    clemens06, bernardi06}, while the residual metallicity trends at
  fixed $\sigma$ in \citet{cooper09} are in the {\it opposite}
  direction of what is observed here for low-$\Delta I_e$ galaxies:
  galaxies in denser environments are {\it older} and slightly {\it
    metal-rich} compared to field galaxies at the same $\sigma$.  The
  \citet{smith08} sample of Coma galaxies show stellar population
  variations at fixed $\sigma$ as a function of cluster-centric
  radius, such that galaxies farther out in the cluster have younger
  ages, higher [Fe/H], and slightly lower [Mg/Fe] than their
  counterparts near the cluster center.  These trends are in the same
  direction as the trends we observe with $\Delta I_e$ and could be
  related.}.  There is observational evidence that this type of
quenching may proceed slowly in massive galaxies \citep{wolf09} and
may not be adequately abrupt to produce the high [Mg/Fe] seen in
high-$\sigma$ galaxies with low $\Delta I_e$.  Furthermore, the
majority of our sample galaxies are central to their haloes; two
thirds of our sample galaxies are identified as the brightest group
galaxies in the \citet{yang07} catalog of SDSS galaxy groups.

The role of satellite quenching as a mechanism for
premature truncation can and should be addressed in future work that
distinguishes between satellite galaxies and those that are central to
their dark matter haloes, as well as studying these trends in $\Delta
I_e$ as a function of environment.

\subsubsection{AGN-driven Quenching}\label{agn_quenching}

Another possible truncation mechanism is very powerful feedback, in
which a substantial quantity of the ISM is heated and removed from the
galaxy potential well before it can form stars.  AGN may be able to
heat and remove substantial fractions of the ISM \citep{granato04,
  scannapieco04, di_matteo05, springel05, hopkins08_etgs}, possibly
with an additional boost to the momentum-driven winds from
AGN-triggered star formation \citep{silk10}.  As an example,
\citet{nesvadba07} present optical and radio observations of two
massive $z = 3.5$ galaxies with strong radio jets and calculate that
the mass-loading of the jet material may represent several times
$10^{10} M_{\odot}$ of ISM.  Such objects may be evidence of powerful
AGN feedback in action.  Interestingly, \citet{hopkins07_bhfp_theory}
find that simulations of black hole growth through major-merger-driven
accrection predict a two-dimensional ``Black Hole Fundamental Plane''
(BHFP), which agrees well with observations
\citep{hopkins07_bhfp_obs}.  However, black hole growth in the
simulations is regulated by the conditions needed for feedback to
power a pressure-driven outflow, such that the resulting black hole
mass depends on the details of the local potential and is surprisingly
insensitive to other merger properties.  In this paradigm, $M_{BH}$
should be unique for a given combination of $\sigma$ and $R_e$.  It is
therefore not obvious how AGN-driven feedback would vary in such a way
as to produce the observed premature truncation in galaxies at fixed
$\sigma$ and $R_e$.

\subsubsection{Supernova-driven Quenching and Differential Halo
  Assembly}\label{sn_quenching} 

Supernova feedback provides another mechanism for injecting energy
into the ISM.  In a model where the SN-driven outflow rate is
comparable to the star formation rate, SN feedback may be relevant for
driving galactic winds and outflows at all mass scales \citep{silk03,
  pipino04}, although many galaxy evolution models require some
further feedback from other sources such as AGN in order to fully
quench star formation in massive galaxies (e.g., \citealt{hatton03,
  croton06, bower06}).  If some galaxies experience stronger SN
feedback for a given star formation rate, for example due to
differences in mass-loading in the winds (e.g., \citealt{pipino09a}),
or if galaxies of a given mass form stars at different rates (e.g.,
due to different accretion histories), the galaxies with stronger
feedback or more rapid star formation may shut down star formation
earlier, resulting in low $\Delta I_e$ galaxies with older ages,
higher [Mg/Fe], and lower total metal enrichment.

Another possibility is that the low $\Delta I_e$ galaxies did not form
the majority of their stars {\it in situ}, but instead assembled
hierarchically at early times from many smaller galaxies.  These small
galaxies might be more susceptible to supernova feedback because they
have shallower potential wells \citep{white78, dekel86, white91,
  benson03}.  This process could shut down star formation at earlier
times and lead to the observed older ages and higher [Mg/Fe] of their
massive descendents, as compared to galaxies at the same $\sigma$ with
higher $\Delta I_e$, which may have formed more of their stars {\it in
  situ} in massive haloes.  A galaxy which formed in many smaller
pieces would also have lower metallicity, both because its ISM
experienced fewer generations of stellar processing resulting in a
lower effective yield, and because SN feedback would be more effective
at removing metal-enriched gas from the galaxy.  

Differences in halo assembly histories are well-modeled by
cosmological simulations.  Thus the new generation of semi-analytic
models, which include predictions for $\sigma$ and $R_e$ based on
merger simulations, may prove important tools for testing
assembly-dependent truncation mechanism.  The successful modeling of
these trends probably requires an improved treatment of satellite
galaxies in the new generation of semi-analytic models (e.g.,
\citealt{guo10}) over previous models, which often fail to correctly
model the stellar populations of satellite galaxies \citep{font08,
  kimm09, pipino09b}.

\subsubsection{Quenching in Massive Dark Matter
  Haloes}\label{halo_quenching} 

Yet another mechanism is quenching by massive haloes (e.g.,
\citealt{birnboim03, keres05, cattaneo08}), when haloes pass over a
critical mass threshold ($M_{crit}$) and shock-heated gas accreting
onto them can no longer cool efficiently.  This process appears to be
able to quench star formation in haloes with $M_{halo} \sim
10^{12}$--$10^{13} M_{\odot}$ \citep{birnboim07}.  When supplemented
with gravitational heating from continued cosmological gas accretion,
this effect may be able to truncate star formation in more massive
haloes as well \citep{dekel08}.  In this scenario, the observed
variations through the FP could be due to stochastic variations in the
halo mass assembly history.  If galaxies with the same $\sigma$ and
$R_e$ today exist in haloes that passed over $M_{crit}$ at different
times, or if mass was accreted differently onto these haloes,
affecting the cooling rate, these galaxies could have different star
formation histories.  The low-$\Delta I_e$ galaxies, which were
truncated earlier, would be those that crossed this critical halo mass
threshold at earlier times.  This scenario should also be
quantitatively testable with semi-analytic models.

\subsubsection{Summary of Truncation Mechanisms}\label{truncation_summary}

These various scenarios for premature truncation make predictions
about how a galaxy's status as a central versus a satellite, the local
environment, central black hole mass, dark matter halo mass, and mass
assembly history should vary at fixed $\sigma$ for galaxies with
different star formation histories and in different parts of FP space.
A combination of further observational programs and improved
semi-analytic models for galaxy formation (that can track values of
$\sigma$ and $R_e$ for galaxies in a cosmological context) should be
able to distinguish between these different truncation mechanisms.
The analysis presented here makes it possible to identify galaxies
that have experienced premature truncation, providing a powerful tool
for exploring the quenching of star formation in galaxies.

\subsection{The Effects of Dry Merging}\label{dry_merging}

The build-up of $L \sim L^*$ red galaxies since $z \sim 1.5$
\citep{bell04, faber07} appears to require a contribution from
dissipationless ``dry'' mergers between galaxies that have already
quenched star formation and occupy the red sequence (e.g.,
\citealt{faber07}).  In this section, we explore the effects of dry
merging on the location of galaxies in FP space and the observed
stellar population trends.  We consider the two extremes of dry
merging: equal-mass major mergers between identical progenitor
galaxies and the build-up of mass through a series of minor mergers
accreting onto a massive central galaxy.  These cases bracket the
range of merger mass ratios.

The scenario of mass-doubling through equal-mass dry mergers has been
explored by \citet{boylan-kolchin05, boylan-kolchin06} using N-body
simulations.  The models use empirically-motivated initial conditions
for the galaxies, combined with \citet{navarro97} dark matter haloes
which are allowed to evolve actively through the course of the
simulation.  More recently, \citet{naab09} have modeled the effects of
mass-doubling through a large number of minor mergers in a
cosmologically-motivated hydrodynamical simulation of a galaxy which
has no mergers with mass ratios greater than 1:8 after $z = 3$.
\citet{naab09} also present a simple analytic argument for the way in
which $\sigma$, $R_e$ and stellar density should scale in mergers,
based on the virial theorem.  We use these studies here to consider
the effects of dry merging on the observed stellar population trends
throughout FP space.

\citet{boylan-kolchin06} find that equal mass dry mergers between
galaxies that lie on the FP result in merger remnants that also lie on
the FP.  The merger remnants have velocity dispersions ($\sigma_f$)
similar to those of the progenitor galaxies ($\sigma_i$), with
$\sigma_f \approx 1.2 \sigma_i$ for the most probable orbital
parameters and $\sigma_f \approx 1.0 \sigma_i$ for radial orbits.  The
remnant effective radii change more, with $R_f \approx 1.6 R_i$ for
the most probably orbital paramters and $R_f \approx 2.5 R_i$ for
radial orbits.  This is in reasonable agreement with the simple
analytic model of \citet{naab09}, who argue that equal-mass dry
mergers should produce $\sigma_f = \sigma_i$ and $R_f = 2 R_i$.

If we assume that the progenitors are identical, lie on the FP, and
have the stellar populations predicted for their location in FP space,
then the resulting merger remnant will have the same stellar
populations as the progenitors, roughly the same $\sigma$, a larger
$R_e$, and will lie on the FP midplane (i.e., it will have $\Delta I_e
= 0$).  This is entirely consistent with the observations in Paper II,
which show that galaxies with the same $\sigma$ but different $R_e$ on
the FP midplane have the same stellar populations.  The virial
scalings of \citet{naab09} also predict almost no motion off the FP
midplane, with $(\Delta I_e)_f = +0.06$ in the merger remnant,
assuming the progenitors had $(\Delta I_e)_i = 0.0$.  Thus equal-mass
dry merging will preserve the stellar population trends identified in
this series of papers and will not tend to move galaxies off the FP
midplane.

Applying the virial scaling arguments of \citet{naab09} to mass
doubling through minor mergers implies $\sigma_f = 0.5 \sigma_i$ and
$R_f = 4 R_i$.  Assuming the progenitors lie on the FP with $(\Delta
I_e)_i = 0.0$, the resulting remnant after many minor mergers should
have $(\Delta I_e)_f = -0.12$.  Thus minor mergers will slowly move
galaxies off the FP toward lower surface brightnesses, while also
lowering $\sigma$ and dramatically increasing $R_e$.  

If the progenitors have the stellar populations inferred from their
location in FP space, the accretion of stars from the lower-$\sigma$
galaxies will tend to lower the SSP age of the galaxy, while also
lowering [Mg/Fe], [Fe/H], and [Mg/H].  However, the hydrodynamical
simulations of \citet{naab09} showed that the majority of stars
accreted from small satellites wind up at large radius.  At $z=0$,
accreted stars account for $\sim 1/7$ of the stars inside 1 kpc,
beyond which the relative contribution of accreted stars slowly
increases.  This means that stellar populations sampled by the SDSS
spectral fibers are dominated by the {\it in situ} component of stars,
not the accreted stars.  Thus the stellar population of the merger
remnant as observed through the SDSS fiber will not change
dramatically from that of the massive progenitor galaxy.

This means that the net effect of a series of minor mergers will be to
move the galaxy to lower $\sigma$ and higher $R_e$, while hardly
changing the central stellar population.  From the virial scaling
argument above, these galaxies will lie slightly off the FP at lower
$\Delta I_e$.  Because the main progenitors started out with higher
$\sigma$, they should have older ages and higher [Mg/Fe] than other
galaxies with the same $\sigma_f$ on the FP midplane.  This is in the
same direction as the observed trends at fixed $\sigma$ to lower
$\Delta I_e$.  However, they should also have higher [Fe/H] and higher
[Mg/H], which is in the opposite direction from the observed trends.
If we relax the aperture assumption and allow the accreted low-mass
galaxies to contribute substantially to the light sampled by the SDSS
fiber, they will tend to contribute stars that are younger with lower
[Mg/Fe], lower [Fe/H], and lower [Mg/H].  This improves the match to
the observed [Fe/H] and [Mg/H] values, but washes out the trends with
age and [Mg/Fe].

All in all, it appears that modest quantities of dry merging will
leave the observed stellar population trends through FP space
essentially unchanged.  Very large numbers of minor mergers may tend
to smear out some of the observed trends.

\section{Conclusions}\label{conclusions}

Papers I and II presented a detailed analysis of the stellar
population properties of quiescent early-type galaxies.  These
population properties were mapped through two familiar early-type
galaxy scaling relations: the color-magnitude relation (Paper I) and
the Fundamental Plane (Paper II).  Paper III supplemented that work
with a detailed study of galaxy mass-to-light ratios, showing that
stellar population mass-to-light ratio effects do not contribute the
majority of the observed variation in $\Delta I_e$.  These works
illustrated that the stellar population properties (and by inference
the star formation histories) of these galaxies form a two-parameter
family.  Furthermore, this two-parameter family maps onto the
large-scale structural properties of the galaxies.

In this work, we have presented what we feel is the clearest and most
intuitive representation of the relevant structural properties for
early-type galaxies: $\sigma$ and $\Delta I_e$.  We have shown how the
various stellar population parameters map onto this structural space,
which corresponds to a cross-section projected through the FP.  We
have also explored in detail the co-variation of mean light-weighted
age, [Fe/H], [Mg/H], and [Mg/Fe] among early-type galaxies.  By
combining these stellar population results with the mass information
from Paper III, we have presented a plausible scenario for early-type
galaxy star formation histories that is consistent with the
multidimensional properties of the data.  Our main conclusions can be
summarized as follows:

\begin{list}{}{}
\item[1.]{The stellar populations of quiescent early-type galaxies
  form a 2D family, similar to the age-metallicity hyperplane of
  \citet{trager00b}.  In the first dimension of the parameter space,
  mean light-weighted age, [Fe/H], [Mg/H], and [Mg/Fe] all increase
  together.  In the second dimension of the parameter space, age and
  [Mg/Fe] increase while [Fe/H] and [Mg/H] decrease.}
\item[2.]{This 2D family of stellar population properties can be
  mapped onto a cross-section projected through the FP, parameterized
  in this work by $\sigma$ and $\Delta I_e$.  The first dimension of
  the parameter space maps onto $\sigma$, such that age, [Fe/H],
  [Mg/H], and [Mg/Fe] all {\it increase} with {\it increasing}
  $\sigma$.  The second dimension maps onto $\Delta I_e$ at fixed
  $\sigma$, such that as $\Delta I_e$ {\it increases}, [Fe/H] and
  [Mg/H] {\it increase} while age and [Mg/Fe] {\it decrease}.}
\item[3.]{Given the systematic behavior described in conclusion 2, it
  is possible to estimate the stellar population properties of a
  quiescent early-type galaxy based on its location in FP space.
  Equations \ref{age_fit}--\ref{mgfe_fit} quantify this mapping.  They
  are shown to reproduce the binned observed stellar population
  properties to roughly within the observational errors.}
\item[4.]{Paper III showed that variations in $\Delta I_e$ are
  primarily due to variations in stellar mass surface density, $\Delta
  \Sigma_{\star,IMF}$.  This means that the observed stellar
  population variations through the thickness of the FP correlate with
  variations in $\Delta \Sigma_{\star,IMF}$, or alternatively with
  variations in $\Delta M_{dyn}/M_{\star,IMF}$.  At fixed $\sigma$,
  galaxies with younger ages, lower [Mg/Fe], and greater total metal
  enrichment (i.e., higher [Fe/H] and [Mg/H]) also have larger
  $\Sigma_{\star,IMF}$ and formed more stars.}
\item[5.]{The observed stellar population variations can be
  interpreted in terms of star formation histories where the start
  time and duration of star formation both depend on $\sigma$ (e.g., a
  ``staged star formation'' model), but in which there are also
  systematic variations in the duration of star formation ($\Delta
  t_{SF}$) at fixed $\sigma$. }
\item[6.]{At fixed $\sigma$, galaxies with longer duration star
  formation have higher $\Sigma_{\star,IMF}$ and higher total
  $M_{\star,IMF}$ at fixed $R_e$.  They also are more enhanced in both
  Fe and Mg, with lower [Mg/Fe] than their counterparts that
  experienced shorter duration star formation.  This strongly suggests
  that they have experienced higher ``conversion efficiencies'' of
  turning their gas into stars.}
\item[7.]{The observed pattern, in which longer duration star
  formation corresponds to the production of more stars, while shorter
  duration star formation produces fewer stars, leads us to propose a
  premature truncation model for quiescent galaxy star formation
  histories.  In this model, star formation proceeds similarly for all
  galaxies of the same $\sigma$ but is truncated at different times.
  Thus some galaxies are shut down while their formation process is
  less complete, leaving them with fewer total stars, lower
  $\Sigma_{\star,IMF}$, and lower metal abundances.}
\end{list}

\acknowledgements

The authors would like to thank Renbin Yan for providing the emission
line measurements used to identify the sample of quiescent early-type
galaxies used here and an anonymous referee whose thoughtful
suggestions improved the quality of this manuscript.  This work was
funded in part by National Science Foundation grant AST 05-07483.
G. G. is supported by a fellowship from the Miller Institute for Basic
Research in Science.  G. G. also acknowledges support from the ARCS
Foundation and from a UCSC Dissertation Year Fellowship.

Funding for the creation and distribution of the SDSS Archive has been
provided by the Alfred P. Sloan Foundation, the Participating
Institutions, the National Aeronautics and Space Administration, the
National Science Foundation, the US Department of Energy, the Japanese
Monbukagakusho, and the Max-Planck Society. The SDSS Web site is
http://www.sdss.org/.

The SDSS is managed by the Astrophysical Research Consortium (ARC) for
the Participating Institutions. The Participating Institutions are the
University of Chicago, Fermilab, the Institute for Advanced Study, the
Japan Participation Group, the Johns Hopkins University, the Korean
Scientist Group, Los Alamos National Laboratory, the
Max-Planck-Institute for Astronomy (MPIA), the Max-Planck-Institute
for Astrophysics (MPA), New Mexico State University, University of
Pittsburgh, University of Portsmouth, Princeton University, the United
States Naval Observatory, and the University of Washington.

\bibliographystyle{apj}
\bibliography{apj-jour,myrefs}

\end{document}